\documentclass[aps,showpacs,nofootinbib,prd,twocolumn,10pt]{revtex4-1}
\usepackage{graphicx}
\usepackage[retainorgcmds]{IEEEtrantools}
\usepackage{amssymb}
\usepackage{amsmath}
\usepackage[svgnames]{xcolor}
\usepackage{mathtools,slashed}
\usepackage{epstopdf}
\usepackage[utf8]{inputenc}
\usepackage{url}
\usepackage[colorlinks,citecolor=DarkGreen,linkcolor=DarkRed,urlcolor=DarkBlue]{hyperref}
\usepackage{subfiles}
\usepackage{physics}
\usepackage[compat=1.1.0]{tikz-feynhand}
\usepackage{tikz}
\usepackage{comment}
\usetikzlibrary{arrows.meta}
\usetikzlibrary{decorations.markings}

\usepackage{newtxtext}
\usepackage{newtxmath}
\usepackage[normalem]{ulem}

\usepackage{dcolumn}
\allowdisplaybreaks

\usepackage{newtxtext}

\usepackage[dvipsnames]{xcolor}

\DeclareMathOperator{\E}{\mathbb{E}}
\DeclareMathOperator{\F}{\mathcal{F}}

\begin{document}
\thispagestyle{empty}

\title{Magnetic field modifications to the pion-quark vertex in the Linear Sigma Model with quarks}

\vspace{50pt}

\author{Alejandro Ayala$^{1,2,3}$}
\author{Fl\'avia Fialho$^2$}
\author{Adri\'an Lara$^1$}
\author{Ana Mizher$^2$}
\author{Javier Rendón$^1$}
\email[Corresponding author: ]{jesus.rendon@correo.nucleares.unam.mx}
\affiliation{
$^1$Instituto de Ciencias
Nucleares, Universidad Nacional Aut\'onoma de M\'exico, Apartado
Postal 70-543, CdMx 04510,
Mexico.\\
$^2$Instituto de Física, Universidade de São Paulo, Rua do Matão, 1371, CEP 05508-090, São Paulo, SP, Brazil.\\
$^3$Instituto de F\' isica Te\'orica, Universidade Estadual Paulista, Rua Dr. Bento Teobaldo Ferraz, 271 - Bloco II, 01140-070 S\~ao Paulo, SP, Brazil.}

\begin{abstract}

We compute the modification to the quark-neutral pion vertex, induced by a constant and uniform magnetic field, using the Linear Sigma Model with quarks as a low energy effective theory of QCD. The vertex is modified even at tree-level since, due to the loss of translational invariance, the calculation should be carried out in configuration space, with the quark described by Ritus wave functions instead of plane waves. We also compute the one-loop vertex modification. These modifications are found for arbitrary Landau levels occupied by the quark. We illustrate the result for the case where the quark occupies the lowest Landau level. To check the result, we also compute the one-loop vertex modification using the Schwinger proper-time method with the quark occupying the lowest Landau level and find the same result as in the case where the Ritus formalism is used.

\end{abstract}

\maketitle

\section{Introduction}
Vertex corrections are an essential component for the refinement and deeper understanding of field theories. Perturbative calculations become indispensable when aiming to achieve a higher level of theoretical precision and consistency with experimental data. Since Schwinger’s seminal work \cite{Schwinger:1948iu}, where the one-loop vertex correction in QED was first computed to extract the vacuum electron anomalous magnetic moment, a systematic framework has been established to calculate perturbatively improved vertices that, together with self-energy calculations, provides a clear interpretation of mass and charge renormalization. 

Strong magnetic fields can significantly modify the vacuum properties of particles, both at tree level and through radiative corrections. Radiative effects arise from the magnetic-field-induced modifications of charged-particle propagators appearing in Feynman loops, whereas tree-level effects originate from the proper description of asymptotic charged states in terms of the Ritus eigenfunctions~\cite{Ritus:1978cj}, instead of the plane waves that characterize the field-free theory. Strong magnetic fields are predicted to be generated in a variety of physical settings, including heavy-ion collisions~\cite{Kharzeev:2007jp,Skokov:2009qp,Voronyuk:2011jd}, astrophysical compact objects such as magnetars~\cite{Duncan:1992hi,Kouveliotou:1998ze}, and the early universe~\cite{Grasso:2000wj}. Strong magnetic fields also provide a powerful probe for studying the properties of the QCD vacuum. For example, at zero temperature, it is well known that magnetic fields catalyze the breaking of chiral symmetry~\cite{Gusynin:1999pq}, even for weak magnetic fields~\cite{Ayala:2006sv}, a phenomenon referred to as magnetic catalysis. However, at nonzero temperatures, magnetic fields suppress condensate formation and lower the critical temperature for chiral symmetry restoration, leading to the phenomenon known as inverse magnetic catalysis~\cite{Bali:2011qj,Bali:2012zg,Bruckmann:2013oba,Farias:2014eca,Ferreira:2014kpa,Ayala:2014gwa,Ayala:2015lta,Ayala:2014iba,Farias:2016gmy,Ayala:2016bbi,Ferrer:2014qka,Ayala:2014uua,Ayala:2018wux,Mueller:2014tea,Mueller:2015fka,Bandyopadhyay:2020zte}.

Modifications of the interaction vertices in weak processes induced by strong magnetic fields have been computed at tree-level for the leptonic decays of charged pions by analyzing their decay widths in Refs.~\cite{Nikishov:1964zza,Nikishov:1964zz,Bali:2018sey,Coppola:2018ygv,Coppola:2019idh,Coppola:2019wvh}. The quark anomalous magnetic moment (AMM) in an extreme magnetic background from perturbative QCD has recently been computed in Ref.~\cite{Fraga:2024klm}. In addition, the one-loop correction to the electron AMM has been improved by taking into account magnetic field corrections in the weak field limit~\cite{Lin:2021bqv}. An earlier result for the electron AMM in the presence of magnetic fields was found in Ref.~\cite{Baier:2000yv}. A more general calculation of the magnetic field induced modifications of the QED vertex, including the analysis of the purely transverse contribution to the fermion AMM for arbitrary Landau levels, has been recently performed in Ref.~\cite{Ayala:2026eja}. Consequences of a magnetic field-dependent quark AMM have been explored in Refs.~\cite{Tavares:2023oln,Farias:2021fci,Fayazbakhsh:2014mca}. More recently, the strong magnetic field modifications of the neutral pion decay width have also been studied in Ref.~\cite{Coppola:2025nus} using the Nambu--Jona-Lasinio model and in Refs.~\cite{Brauner:2017uiu,Adhikari:2024vhs} using chiral perturbation theory and  in Refs.~\cite{Ayala:2020muk,Ayala:2020dxs}, approximations to the magnetic corrections of the boson self-coupling and boson-fermion coupling in the linear sigma model with quarks (LSMq).

Motivated by these developments, here we compute both the general expressions for the tree-level and one-loop magnetic-field-induced corrections to the quark-neutral pion-quark vertex, within the LSMq. We present explicit results for the case of the $u\pi^0 u$ case. These corrections are relevant for describing scattering processes in a magnetized medium, for systems undergoing spontaneous chiral symmetry restoration, and for decay processes  influenced  by the presence of an external magnetic field. The work is structured as follows: In Sec.~\ref{sec: the model}, we summarize the ingredients of the LSMq. In Sec.~\ref{sec: proper time}, we discuss the form of the propagators for the charged particles of the model, both in the Ritus and Schwinger's proper time formalisms. In Sec.~\ref{sec: vertex}, we use the Ritus formalism to perform the calculations that correspond to the quark-neutral pion vertex correction for an arbitrary strength of the magnetic field, at tree-level in Sec.~\ref{vertex-tree}, and at one-loop level in Sec.~\ref{vertex-loop}, for arbitrary Landau levels occupied by the quark. We illustrate the calculation, showing the result explicitly in the lowest Landau level (LLL). In Sec.~\ref{secV} we perform the vertex correction at one-loop level, starting explicitly from the quark in the LLL using the Ritus formalism. In Sec.~\ref{Schwing}, we independently evaluate the same one-loop correction in the LLL using Schwinger's proper-time formalism and demonstrate the equivalence between both approaches. In Sec.~\ref{results} we present the numerical results for the case when the quark occupies the lowest Landau level (LLL). We finally summarize and conclude in Sec.~\ref{conclusions} and leave for the appendices the explicit computation of the integrals involved, as well as the derivation of some of the elements required in the calculation.

\section{Linear Sigma Model with quarks}\label{sec: the model}
The $SU(2)\times SU(2)$ LSMq is an effective theory of QCD that includes quarks together with the lightest scalar and pseudoscalar mesons as explicit degrees of freedom. It incorporates chiral symmetry breaking and is particularly well-suited for studying hadronic physics in the light-meson sector. The Lagrangian is given by
\begin{eqnarray}
\mathcal{L}&=&\frac{1}{2}(\partial_{\mu}\sigma)^{2}+\frac{1}{2}(\partial_{\mu}\vec{\pi})^{2}+\frac{a^{2}}{2}(\sigma^{2}+\vec{\pi}^{2})-\frac{\lambda}{4}(\sigma^{2}
+\vec{\pi}^{2})^{2}\nonumber\\
&+&i\bar{\psi}\gamma^{\mu}\partial_{\mu}\psi-ig\gamma^{5}\bar{\psi} \vec{\tau} \cdot \vec{\pi}\psi-g\bar{\psi}\psi\sigma\,,
\label{lagrangian}
\end{eqnarray}
where $\vec{\pi}=(\pi_1,\pi_2,\pi_3)$ represents an isospin triplet, $\psi=(u,d)$ represents an isospin doublet, and the $\sigma$ scalar is an isospin singlet. The parameters $\lambda$ and $g$ correspond to the boson self-coupling and the boson-fermion coupling, respectively.\\
\indent After spontaneous symmetry breaking, the $\sigma$ field becomes
\begin{equation}
    \sigma\rightarrow\sigma+v\,.
\end{equation}
As a consequence of this shift, the Lagrangian $\mathcal{L}$ reads
\begin{eqnarray}
    \mathcal{L}&=&\frac{1}{2}\partial_{\mu}\sigma \partial^{\mu}\sigma+\frac{1}{2}\partial_{\mu}\pi_{0}\partial^{\mu}\pi_{0}+\partial_{\mu}\pi_{-}\partial^{\mu}\pi_{+}\nonumber\\
    &-&\frac{1}{2}m_{\sigma}^{2}\sigma^{2}-\frac{1}{2}m_{0}^{2}\pi_{0}^{2}-m_{0}^{2}\pi_{-}\pi_{+}+i\bar{\psi}\slashed{\partial}\psi\nonumber\\
    &-&m_{f}\bar{\psi}\psi+\frac{a^2}{2}v^2-\frac{\lambda}{4}v^4\ +\ hv+\mathcal{L}_{int},
    \label{linearsigmamodelSSB}
\end{eqnarray}
where $hv$ represents an explicit symmetry breaking term to give the pions a finite mass. The charged-pion fields are defined as
\begin{equation}
 \pi_\pm=\frac{1}{\sqrt{2}}(\pi_1\mp i\pi_2),
\end{equation}
and the interaction Lagrangian takes the form
\begin{equation}
\begin{split}
    \mathcal{L}_{int}&=-\frac{\lambda}{4}\sigma^{4}-\lambda v\sigma^{3}-\lambda v^{3}\sigma-\lambda\sigma^{2}\pi_{-}\pi_{+} -2\lambda v \sigma\pi_{-}\pi_{+}\\
    &-\frac{\lambda}{2}\sigma^{2}\pi_{0}^{2}-\lambda v\sigma \pi_{0}^{2}-\lambda \pi_{-}^{2}\pi_{+}^{2}-\lambda\pi_{-}\pi_{+}\pi_{0}^{2}-\frac{\lambda}{4}\pi_{0}^{4}\\ 
    &+a^{2}v\sigma -g\bar{\psi}\psi\sigma-ig\gamma^{5}\bar{\psi}\left(\tau_{+}\pi_{+}+\tau_{-}\pi_{-}+\tau_{3}\pi_{0}\right)\psi.
    \label{interactinglagrangian}
\end{split}    
\end{equation}
The masses appearing in Eq.~(\ref{linearsigmamodelSSB}) are explicitly given by
\begin{align}
     m_{\sigma}^{2}&=3\lambda v^2-a^2, \nonumber \\
     m_{0}^{2}&=\lambda v^2-a^2, \nonumber \\ 
     m_{f}&=gv.
\label{masses}
\end{align}
We introduce the effect of an external magnetic field $\mathbf{B}$ through a covariant derivative in the Lagrangian density in Eq.~(\ref{linearsigmamodelSSB}), that is
\begin{equation}
 \partial_\mu\to D_\mu=\partial_\mu+ieA_\mu,
\end{equation}
where $A^{\mu}$ is the vector potential that couples a charged particle with the external magnetic field. In this work, we take this field to be uniform in space and constant in time, choosing the $z$-axis as its direction.

\section{Magnetic field-dependent propagators}\label{sec: proper time}
The charged particle propagators are modified in the presence of a background magnetic field. In this section, we summarize the explicit expressions for the scalar and fermion propagators immersed in the magnetic field. Their representations are provided both in the Ritus basis and in the Schwinger proper-time formalism.

We start by discussing the Schwinger proper-time formalism, and then discuss the Ritus representation. Although not directly used in this work, for completeness, we first present  the scalar charged propagator, which can be written as
\begin{equation}
D_{b}(x,x')=e^{i\Phi_P(x,x')}D_{b}(x-x')\,,
\label{scalarprop}
\end{equation}
where the Schwinger phase $\Phi_P(x,x')$ is defined as
\begin{eqnarray}
\Phi_P(x,x')=Q_{P}\int_x^{x'}d\xi_\mu \left[
A^\mu(\xi) + \frac{1}{2}F^{\mu\nu}(\xi-x')_\nu
\right],
\label{phase}
\end{eqnarray}
and $Q_{P}$ denotes the electric charge of a particle $P$. The translationally and gauge invariant part of the propagator $D_{b}(x-x')$ can be written in terms of its Fourier transform as

\begin{equation}
    D_{b}(x-x')=\int \frac{d^{4}p}{(2\pi)^{4}}D_{b}(p)e^{-ip\cdot(x-x')}
\end{equation}
where
\begin{eqnarray}
iD_{b}(p)&=&\int_0^\infty \frac{ds}{\cos(|q_bB|s)}e^{is\left(p_\parallel^2-p_\perp^2\frac{\tan(|q_bB|s)}{|q_bB|s}-m_b^2+i\epsilon \right)}\,.\nonumber\\
\label{bosonpropagatormomentumspace}
\end{eqnarray}
The scalar propagator can also be written in terms of Ritus functions as
\begin{equation}
     D_b(x,y) = \sumint_{\bar{q}}\F^{Q}(x,\bar{q})D_b(k,q_{\parallel})\bar{\F^{Q}}(y,\bar{q})
     \label{scalar_propagator}
\end{equation}
where in the Landau gauge 2 (LG2) we have
\begin{equation}
     \sumint_{\bar{q}}=\frac{1}{2\pi}\int \frac{dq_{0}}{2\pi}\frac{dq_{2}}{2\pi}\frac{dq_{3}}{2\pi}\,,
\end{equation}
and
\begin{equation}
    D_b(k,q_{\parallel})=\frac{1}{q_{\parallel}^2 - m_b^2-(2k+1)B_Q +i\epsilon}
\end{equation}
with $B_Q=|qB|$. The Ritus functions $\F^{Q}(x,\bar{q})$ that appear in Eq.~({\ref{scalar_propagator}}) can be explicitly written in the LG2 as
\begin{eqnarray}
     \F^{Q}(x,\bar{q}) &=& N_k e^{-i(q_0x_0 - q_2x_2-x_3x_3)}\nonumber\\
     &\times&D_k\left(\sqrt{2B_Q}\left(x_1 - \frac{sq_2}{B_Q}\right)\right)\,,
     \label{Ritus_Function}
\end{eqnarray}
where 
\begin{eqnarray}
N_k &=& (4\pi B_Q)^{1/4}/\sqrt{k!},\nonumber\\
D_k(x) &=& 2^{-k/2}e^{-x^2/4}H_k(x/\sqrt{2}),
\label{defND}
\end{eqnarray}
with $H_k$ the Hermite polynomial of degree $k$. We also use the definition $\bar{q}=(q_0,k,q_2,q_3)$.

We now present the corresponding results for the fermion propagator. Once again, we begin with Schwinger's proper-time formalism. The fermion propagator can be written as
\begin{equation}
    S_f(x,x')=e^{i\Phi_P(x,x')}S_f(x-x'),
    \label{fermionpropagatorincoordinatespace}
\end{equation}
where the translationally and gauge invariant part $S_f(x-x')$ is given by
\begin{equation}
    S_f(x-x')=\int \frac{d^{4}p}{(2\pi )^{4}}S_f(p)e^{-ip\cdot(x-x')}, \label{Fouriertransformfermionpropagator}
\end{equation}
with
\begin{eqnarray}
    iS_f(p)&=&\int_0^\infty \frac{ds}{\cos(|q_fB|s)}e^{is\left(p_\parallel^2-p_\perp^2\frac{\tan(|q_fB|s)}{|q_fB|s}-m_f^2+i\epsilon\right)}\nonumber\\
&\times&\left[
\Big(
\cos(|q_fB|s) + \gamma_1\gamma_2\sin(|q_fB|s)\text{sign}(q_fB)
\Big)\right.\nonumber\\
&\times&\left.\left(m_f +\slashed{p}_\parallel\right) - \frac{\slashed{p}_\perp}{\cos(|q_fB|s)}
\right].\label{fermionpropagatormomentumspace}
\end{eqnarray}
In terms of Ritus functions we can also write the fermion propagator as
\begin{equation}
    S_f(x,y) = \sumint_{\bar{q}}\E^{Q}(x,\bar{q})S_f(k,q_{\parallel})\bar{\E}^{Q}(y,\bar{q})
    \label{fermion_propagator}
\end{equation}
where
\begin{equation}
S_f(k,q_{\parallel})=\frac{\slashed{\Pi}_s + m_f}{q_{\parallel}^2 - m_f^2-2kB_Q}\,,
\label{Ritusprop}
\end{equation}
with $\Pi_s^\mu = (q_0,0,-s\sqrt{2kB_Q},q_3)$.\\
The Ritus functions $\E^{Q}(x,\bar{q})$ appearing in Eq.~(\ref{fermion_propagator}) are given by
\begin{equation}
    \E^{Q}(x,\bar{q}) = \sum_{\lambda = \pm} \Delta^\lambda\F_{Q}(x,\bar{q}_\lambda) 
    \label{Ritus_11}
\end{equation}
and
\begin{equation}
    {\bar{\E}^{Q}}(y,\bar{q})=\gamma^{0}{\E}^{Q}(y,\bar{q})^{\dagger}\gamma^{0}
    \label{barE}
\end{equation}
with $\Delta^{\lambda} = (1 + i\lambda\gamma^1 \gamma^2)/2$. In the LG2 $\overline{q}_\lambda = (q_0,k_{s\lambda},q_2,q_3)$ with $k_{s\lambda} = k - (1-s\lambda)/2$,
with the function $\F_Q(x,\bar{q})$, given by Eq.~(\ref{Ritus_Function}).

\section{Vertex corrections}\label{sec: vertex}
In this section, we analyze the vertex modifications induced by the magnetic field for the neutral pion and a quark, that for definitiveness, we take as $u$. We perform the calculation both at tree- and at one-loop level, following closely the notation of Refs.~\cite{GomezDumm:2023owj,Coppola:2018ygv}.
\begin{figure}[t]
    \centering
    \includegraphics[width=0.6\linewidth]{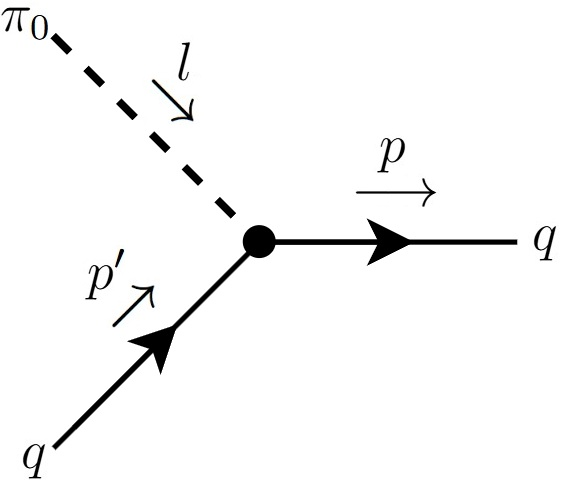}
    \caption{Feynman diagram corresponding to the tree-level magnetic modification of the $\pi_{0} qq$ vertex, represented by the blob joining the $\pi_0$ and the $q$ lines.}
    \label{Tree}
\end{figure}

\subsection{TREE LEVEL VERTEX CORRECTION}\label{vertex-tree}
The amplitude for the process $u+\pi_{0}\rightarrow u$ at tree-level which corresponds to the Feynman diagram of Fig.~\ref{Tree} is given by

\begin{equation}
    \mathcal{V} =\langle{u}|\ i\int d^{4}x\mathcal{L}_{int}(x)|u\pi_{0}\rangle\,, 
    \label{amplitude_tree}
\end{equation}

for which the interaction Lagrangian $\mathcal{L}_{int}(x)$ is,
\begin{equation}
   \mathcal{L}_{int}(x) = -ig\bar{u}\gamma^5u\pi_0\,.
   \label{L_int}
\end{equation}

The expansion for the quark field in terms of creation and annihilation operators has the form
\begin{eqnarray}
    \!\!\!\!\!\!\!\psi(x) &\!\!\!=\!\!\!\!\!& \sumint_{\{\bar{q}E_Q\}}\frac{1}{2E_Q}\nonumber\\
    &\!\!\!\times\!\!\!&\sum_{a = 1,2} \left\{  b(\breve{q},a)U(x,\bar{q},a) + d^{\dagger}(\breve{q},a) V(x,\bar{q},a)  \right\},
    \label{quark_field}
\end{eqnarray}
where the quark energy is given by 
\begin{eqnarray}
E_Q = \sqrt{m^2 + 2kB_Q + (q_3)^2},   
\end{eqnarray}
with $k \ge 0$, and $\breve{q}=(k,q_2,q_3)$, also we have used the shorthand notation
\begin{equation}
    \sumint_{\{\bar{q}E_Q\}} = \sumint_{\bar{q}}2\pi\delta(q_0 - E_Q )\,.
\end{equation}
This expression depends on the choice of gauge and working in the LG2, it explicitly looks like
\begin{equation}
      \sumint_{\bar{q}} \equiv \frac{1}{2\pi}\sum_{k = 0}^{\infty}\int \frac{dq_0}{2\pi}\frac{dq_2}{2\pi}\frac{dq_3}{2\pi}\,.
\end{equation}
The corresponding expansion for the neutral pion field in terms of creation and annihilation operators is
\begin{equation}
    \pi_0(x) = \int \frac{d^4p}{(2\pi)^4}\frac{2\pi \delta(p_0 - E_\pi)}{2E_\pi}(a_p e^{-ip\cdot x} + a_p^{\dagger}e^{ip\cdot x})\,,
    \label{pion_field}
\end{equation}
where the pion energy is given by $E_\pi = \sqrt{|\Vec{p}|^2 + m^{2}_{\pi}}$.\\

Canonical quantization imposes the following anti-commutation relations for the quark fields
\begin{eqnarray}
    \{b(\breve{q},a),b(\breve{q}',a')\} &=& \{d(\breve{q},a),d(\breve{q}',a')\} = 0,\nonumber \\
    \{b(\breve{q},a),d(\breve{q}',a')\} &=& \{b(\breve{q},a),d(\breve{q}',a')^{\dagger}\} = 0,\nonumber\\
    \{b(\breve{q},a),b(\breve{q}',a')^{\dagger}\} &=& 2E_{Q}(2\pi)^3\delta_{\chi\chi'}\delta_{aa'}\delta_{kk'}\delta(q_3-{q'}_{3})\nonumber\\ \{d(\breve{q},a),d(\breve{q}',a')\}^{\dagger} &=& 
    2E_{Q}(2\pi)^3\delta_{\chi\chi'}\delta_{aa'}\delta_{kk'}\delta(q_3-{q'}_{3}),\nonumber\\
\label{quark_commutation}
\end{eqnarray}
and the commutation relations for the pion field
\begin{equation}
\begin{gathered}
    [a_p,a_{p'}] = [a_p^{\dagger},a_{p'}^{\dagger}] = 0\,, \\
    [a_p,a_{p'}^{\dagger}] = 2E_{\pi}(2\pi)^3\delta^{(3)}(\Vec{p} - \Vec{p}')\,.
\end{gathered}
\label{pion_commutation}
\end{equation}
Notice that the quark field expansion in Eq.~(\ref{quark_field}) is expressed in terms of Ritus eigenfunctions, as these correspond to charged states, whereas the pion field expansion in Eq.~(\ref{pion_field}) is written in terms of plane waves, reflecting its neutral nature. The commutation relations in Eq.~(\ref{pion_commutation}) are the standard ones. In contrast, the anti-commutation relations in Eq.~(\ref{quark_commutation}) explicitly depend on the choice of gauge, reflecting once again the charged nature of the corresponding particles.

Inserting the interaction Lagrangian together with the expansion of the fields into Eq.~(\ref{amplitude_tree}), and using the commutation and anti-commutation rules, we obtain the following expression for the tree-level vertex
\begin{equation}
    \mathcal{V}^{\mbox{\small{ (tree)}}} = -g \int d^{4}x\,\bar{U}(x,\bar{p}',a')\gamma^5{U}(x,\bar{p},a)e^{-il\cdot x}\,,
    \label{amplitude}
\end{equation}
where the explicit expression for $U(x,\bar{p},a)$
is given by

\begin{equation}
\begin{gathered}
    U(x,\bar{p},a) = \E^{Q}(x,\bar{p})u_Q(k,x_3,a), 
\end{gathered}
\label{spinors}
\end{equation}

and the spinor $u_Q(k,x_3,a)$
is written as
\begin{equation}
    u_Q(k,x_3,a)\!=\!\frac{\left[\slashed{\Pi}_s(E_Q,k,x_3)\!+\!m_f\mathcal{I} \right]}{\sqrt{2(E_Q + m)}}
    \begin{pmatrix}
        \phi^{(a)}\\
        \phi^{(a)}
    \end{pmatrix}.
\end{equation}
For completeness, we also write the expression for the negative energy solution
\begin{equation}
V(x,\bar{p}',a) = \Tilde{\E}^{-Q}(x,\bar{p}')v_{-Q}(k,x_3,a)\,.
\end{equation}
where

\begin{equation}
   v\!_{-Q}(k,\!x_3,\!a)\!= \!\frac{\left[\!- \slashed{\Pi}_{\!-s}(E_Q,k,x_3)\! +\!m_f\mathcal{I} \right]}{\sqrt{2(E_Q\!+\!m)}}
   \begin{pmatrix}
      \!\! \Tilde{\phi}^{(a)}\\
       - \Tilde{\phi}^{(a)}\!\!
    \end{pmatrix},
\end{equation}

\begin{figure}[t!]
    \centering
    \includegraphics[width=0.7\linewidth]{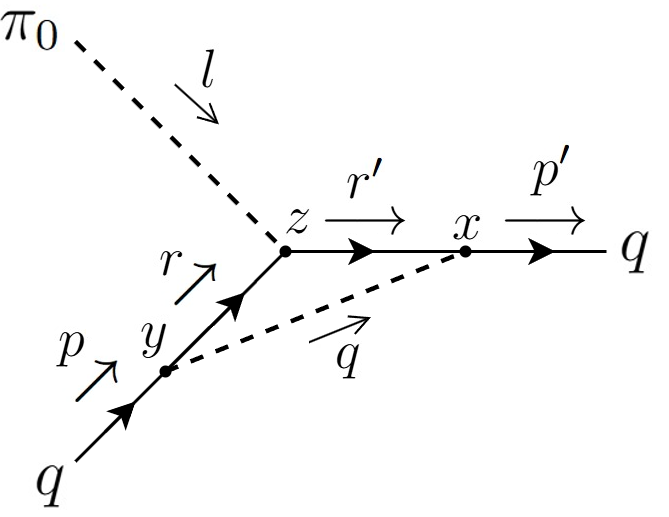}
    \caption{Feynman diagram corresponding to the one-loop magnetic modification of the $\pi_{0} q\bar{q}$ vertex.}
    \label{Loop}
\end{figure}

with $\phi^{(1)\dagger} = - \Tilde{\phi}^{(2)\dagger} = (1,0)$, $\phi^{(2)\dagger} = \Tilde{\phi}^{(1)\dagger} = (0,1)$ and $\Pi_s^\mu(q_0,k,q_3) = (q_0,0,-s\sqrt{2kB_Q},q_3) $.
The Weyl representation for the Dirac matrices is used in this work,
\begin{equation}
    \gamma_0 = 
    \begin{pmatrix}
        0 & \mathcal{I}\\
        \mathcal{I} & 0\\
    \end{pmatrix}\qquad \Vec{\gamma} = 
    \begin{pmatrix}
        0 & \Vec{\sigma}\\
        - \Vec{\sigma} & 0\\
    \end{pmatrix}.
\end{equation}
Notice that  Eq.~(\ref{spinors}) contains the Ritus function $\E^{Q}(x,\bar{p})$, previously defined in Eq.~(\ref{Ritus_11}). For completeness, we also write the function $\Tilde{\E}^{-Q}(x,\bar{p}')$, given by~\cite{Ritus:1978cj}
   \label{Ritus_1}

\begin{equation}
    \Tilde{\E}^{-Q}(x,\bar{p}') = \sum_{\lambda = \pm} \Delta^\lambda\F^{*}_{-Q}(x,\bar{p}'_{-\lambda}),
\label{Ritus_2}
\end{equation}
with the function $\F_Q(x,\bar{q})$ defined in Eq.~(\ref{Ritus_Function}). 
Inserting the explicit forms for $\bar{U}(x,\bar{p},a)$ and ${U}(x,\bar{p}',a')$ in Eq.~(\ref{amplitude}) we can write the vertex $\mathcal{V}^{\mbox{\small{(tree)}}}$ as
\begin{equation}
    \mathcal{V}^{\mbox{\small{ (tree)}}} = -g\bar{u}_Q \xi(\bar{p},\bar{p}',l)u_{Q}\,,
\end{equation}
where
\begin{eqnarray}
    \xi(\bar{p},\!\bar{p}'\!,\!l)\! &=& \!\gamma^5\!\sum_{\lambda =\pm}\!\Delta^{\lambda}(2\pi)^3\delta^2(p'_\parallel\!-p_\parallel\!-\! l_\parallel)\delta(p'_2 \!- \!p_2 \!-\! l_2)\nonumber\\
    &\times&\mathcal{J}(k_{s\lambda},k'_{s\lambda}),
    \label{formfactortree}
\end{eqnarray}
with
\begin{equation}
    \mathcal{J}(k_{s\lambda},k'_{s\lambda}) = e^{is\frac{l^1(p'^2+p^2)}{2B_Q}}\mathcal{G}_{k_{s\lambda},k'_{s\lambda}}(l_\perp)\,,
    \label{J}
\end{equation}
and
\begin{eqnarray}
    \mathcal{G}_{k_{s\lambda},k'_{s\lambda}}(l_\perp) &\!\! = \!\!&2\pi i^{k_{s\lambda} -k'_{s\lambda}} \sqrt{\frac{k'_{s\lambda}!}{k_{s\lambda}!}}\left(\frac{l_{\perp}^2}{2B_Q}\right)^{\frac{k_{s\lambda} -k'_{s\lambda}}{2}}\nonumber\\
    &\!\!\times\!\!&e^{-\frac{l^2_{\perp}}{4|B_Q|}}e^{is\phi_\perp(k_{s\lambda} -k'_{s\lambda})}\nonumber \\
    &\!\!\times\!\!&  L_{k'_{s\lambda}}^{k_{s\lambda} -k'_{s\lambda}}\left(\frac{l_{\perp}^2}{2B_Q}\right). 
    \label{G}
\end{eqnarray}
Equation~(\ref{formfactortree}), together with the functions defined in Eqs.~(\ref{J}) and~(\ref{G}), represents the tree-level modification to the point-like $\pi_0 qq$. Since, in the presence of the magnetic field, overall momentum is not conserved, the modification is written as a Gaussian form factor in the transverse direction, modulated by a Laguerre polynomial that depends on the quark Landau levels. The modified vertex also depends on the polarization state of the quark pair. We thus see that the vertex is modified even at tree-level. However, in order to get a more precise perturbative picture of the vertex modifications induced by the magnetic field, we now proceed to compute the one-loop correction.

\begin{widetext}
\subsection{One-loop vertex}\label{vertex-loop}
The Feynman diagram representing the one-loop $\pi_{0}qq$ vertex is shown in Fig.~\ref{Loop}. Its explicit expression is given by
\begin{equation}
    \mathcal{V}^{\mbox{\small{ (1)}}} =i^3\int d^4xd^4yd^4z \bar{U}(p',x)(-ig\gamma^5)S_f(x,z)(-ig\gamma^5)S_f(z,y)(-ig\gamma^5)D(y,x)U(y,p)e^{-il\cdot z}.
    \label{vertex}
\end{equation}
Using Eqs.~(\ref{fermion_propagator}),~(\ref{Ritusprop}) and~(\ref{spinors}) into Eq.~(\ref{vertex}) we can write $\mathcal{V}^{\mbox{\small{(1)}}}$ as
\begin{equation}
    \begin{aligned}
        \mathcal{V}^{\mbox{\small{ (1)}}} =ig^3\int d^4xd^4yd^4z \sumint_{\bar{r}} \sumint_{\bar{r'}}\int \frac{d^4q}{(2\pi)^4} \frac{e^{-iq\cdot (y-x)}e^{-il\cdot z}}{(r_{\parallel}^2 -m^2 -2k_1B_Q)(r_{\parallel}'^2 -m^2 -2k_2B_Q)(q^2-m_\pi^2)}\\
        \bar{u}(k',p',a')\bar{\E}^{Q}(x,\bar{p'})\E^{Q}(x,\bar{r'})(\slashed{\Pi}_s(\bar{r'}) - m)\gamma^5\bar{\E}^{Q}(z,\bar{r'})\E^{Q}(z,\bar{r})(\slashed{\Pi}_s(\bar{r}) - m)\bar{\E}^{Q}(y,\bar{r})\bar{\E}^{Q}(y,\bar{p})u(k,p,a)
    \end{aligned}
\end{equation}
It is convenient to define the following integrals 
\begin{eqnarray}
    I_x &=& \int d^4x \, \bar{\E}^{Q}(x,\bar{p}')\E^{Q}(x,\bar{r}')e^{iq\cdot x},\\
    I_y &=& \int d^4y \, \bar{\E}^{Q}(y,\bar{r})\E^{Q}(y,\bar{p})e^{-iq\cdot y},\\
    I_z &=& \int d^4z \, \bar{\E}^{Q}(z,\bar{r}')\E^{Q}(z,\bar{r})e^{-il\cdot z}.
\end{eqnarray}
\end{widetext}
These three spatial integrals, together with the integral involved in the tree-level calculation, have the same structure. They are explicitly computed in Appendix~\ref{App: Spatial_Integrals}. The result is
\begin{eqnarray}
    I_x \!&=& \!\sum_{\lambda_1=\pm}\!\Delta^{\lambda_1}(2\pi)^3\delta^2(p'_\parallel\!-\!r'_\parallel\!+q_\parallel)\nonumber\\
    &\times&\delta(p'_2\!-\!r'_2\!+q_2)\mathcal{J}_{\,1\,k'_{s\lambda_1},k_{2s\lambda_1}}(p'_\perp,r'_\perp,q_\perp),\\
    I_y\!&=&\!\sum_{\lambda_2=\pm}\!\Delta^{\lambda_2}(2\pi)^3\delta^2(r_\parallel\!-\!p_\parallel\!-\!q_\parallel)\nonumber\\
    &\times&\delta(r_2\!-\!p_2\!-\!q_2)\mathcal{J}_{\,2\,k_{1s\lambda_2},k_{s\lambda_2}}(r_\perp,p_\perp,q_\perp),\\
    I_z\!&=& \!\sum_{\lambda_3=\pm}\!\Delta^{\lambda_3}(2\pi)^3\delta(r'_\parallel\!-\!r_\parallel\!-\!l_\parallel)\nonumber\\
    &\times&\delta(r'_2\!-\!r_2\!-\!q_2)\mathcal{J}_{\,3\,k_{2s\lambda_3},k_{1s\lambda_3}}(r'_\perp,r_\perp,l_\perp),
\end{eqnarray}
where
\begin{widetext}
\begin{eqnarray}
    \mathcal{J}_{\,1}(k'_{s\lambda_1},k_{2s\lambda_1}) := \mathcal{J}_{\,1\,k'_{s\lambda_1},k_{2s\lambda_1}}(p'_\perp,r'_\perp,q_\perp) = e^{-\frac{i s l_1(p_2 + r_2)}{2B_Q}}G_{k'_{s\lambda_1},k_{2s\lambda_1}}(l_\perp)\\
    \mathcal{J}_{\,2}(k_{1s\lambda_2},k_{s\lambda_2}) := \mathcal{J}_{\,2\,k_{1s\lambda_2},k_{s\lambda_2}}(r_\perp,p_\perp,q_\perp) = e^{\frac{i s l_1(r'_2 + p'_2)}{2B_Q}}G_{k_{1s\lambda_2},k_{s\lambda_2}}(l_\perp)\\
    \mathcal{J}_{\,3}(k_{2s\lambda_3},k_{1s\lambda_3}):=\mathcal{J}_{\,3\,k_{2s\lambda_3},k_{1s\lambda_3}}(r'_\perp,r_\perp,l_\perp) = e^{\frac{i s q_1(r_2 + r'_2)}{2B_Q}}G_{k_{2s\lambda_3},k_{1s\lambda_3}}(q_\perp).
\end{eqnarray}
Therefore, the expression for $\mathcal{V}^{\mbox{\small{(1)}}}$ becomes
\begin{equation}
    \begin{split}
        \mathcal{V}^{\mbox{\small{ (1)}}} =&ig^3 \sumint_{\bar{r}}\sumint_{\bar{r'}}\int \frac{d^4q}{(2\pi)^4} \frac{1}{(r_{\parallel}^2 -m^2 -2k_1B_Q +i\epsilon)(r_{\parallel}'^2 -m^2 -2k_2B_Q +i\epsilon)(q^2-m_\pi^2 +i\epsilon)}\\ 
        &\times\sum_{\lambda_1,\lambda_2,\lambda_3=\pm}\bar{u}(k',p',a') \Delta^{\lambda_1}(\slashed{\Pi}_s(\bar{r}') - m)\gamma^5  \Delta^{\lambda_3}(\slashed{\Pi}_s(\bar{r}) - m) \Delta^{\lambda_2}u(k,p,a)\\
       & \times(2\pi)^9 \delta^2(p'_\parallel-r'_\parallel+q_\parallel)\delta^2(r_\parallel-p_\parallel-q_\parallel)\delta^2(r_\parallel-r'_\parallel-l_\parallel)\\
&\times\delta(p'_2-r'_2+q_2)\delta(r_2-p_2-q_2)\delta(r_2-r'_2-l_2)\\
&\times\mathcal{J}_{\,1}(k'_{s\lambda_1},k_{2s\lambda_1})\mathcal{J}_{\,2}(k_{1s\lambda_2},k_{s\lambda_2})\mathcal{J}_{\,3}(k_{2s\lambda_3},k_{1s\lambda_3}) \,.
    \end{split}.
\end{equation}

Integrating over $\vec{r}$ and $\vec{r}'$, and with $\Pi_{s\perp}(k) = (0,-s\sqrt{2kB_Q})$ , we get
\begin{equation}
    \begin{split}
        \mathcal{V}^{\mbox{\small{ (1)}}} =& ig^3 \frac{1}{(2\pi)^2}\sum_{k_1,k_2 = 0}^{\infty}\int \frac{d^4q}{(2\pi)^4} \frac{\mathcal{J}_{\,1}(k'_{s\lambda_1},k_{2s\lambda_1})\mathcal{J}_{\,2}(k_{1s\lambda_2},k_{s\lambda_2})\mathcal{J}_{\,3}(k_{2s\lambda_3},k_{1s\lambda_3})}{(r_{\parallel}^2 -m^2 -2k_1B_Q +i\epsilon)(r_{\parallel}'^2 -m^2 -2k_2B_Q +i\epsilon)(q^2-m_\pi^2 +i\epsilon)}\\ 
        &\times\sum_{\lambda_1,\lambda_2,\lambda_3=\pm}\bar{u}(k',p',a') \Delta^{\lambda_1}(\slashed{p}'_{\parallel} + \slashed{q}_\parallel + \slashed{\Pi}_{s\perp}(k_2) - m)\gamma^5  \Delta^{\lambda_3}(\slashed{p}_{\parallel} + \slashed{q}_\parallel + \slashed{\Pi}_{s\perp}(k_1) - m) \Delta^{\lambda_2}u(k,p,a)\\
       & \times(2\pi)^3 \delta^2(p'_\parallel-p_\parallel-l_\parallel)\delta(p_2-p'_2-l_2).
    \end{split}
\end{equation}
Performing the spinor algebra, and using the Dirac equation $\bar{u}(k',p',a')(\slashed{\Pi}_{s}(\bar{p}') - m) = 0$ to write expressions involving parallel momenta in favor of perpendicular momenta, we obtain
\begin{eqnarray}
    \label{dgamma}
        \mathcal{V}^{\mbox{\small{ (1)}}} &=& ig^3 \frac{1}{(2\pi)^2}\sum_{k_1,k_2 = 0}^{\infty}\int \frac{d^4q}{(2\pi)^4} \frac{(2\pi)^3 \delta^2(p'_\parallel-p_\parallel-l_\parallel)\delta(p_2-p'_2-l_2)}{((p+q)_{\parallel}^2 -m^2 -2k_1B_Q +i\epsilon)((p'+q)_{\parallel}^2 -m^2 -2k_2B_Q +i\epsilon)(q^2-m_\pi^2 +i\epsilon)}\nonumber
        \\ 
        &\times&\sum_{\lambda_1=\pm}\bar{u}(k',p',a')\gamma^5\Delta^{\lambda_1}\nonumber\\
        &\times&\left\{\mathcal{J}_{\,1}(k'_{s\lambda_1},k_{2s\lambda_1})\mathcal{J}_{\,2}(k_{1s,-\lambda_1},k_{s,-\lambda_1})\mathcal{J}_{\,3}(k_{2s,-\lambda_1},k_{1s,-\lambda_1})\left.\left[(\slashed{\Pi}_{s\perp}(k') + \slashed{\Pi}_{s\perp}(k_2))\slashed{q}_\parallel -(\slashed{\Pi}_{s\perp}(k')+\slashed{\Pi}_{s\perp}(k_2))\slashed{\Pi}_{s\perp}(k)\right]\right.\right.\nonumber\\        
        &-& \mathcal{J}_{\,1}(k'_{s\lambda_1},k_{2s\lambda_1})\mathcal{J}_{\,2}(k_{1s\lambda_1},k_{s\lambda_1})\mathcal{J}_{\,3}(k_{2s-\lambda_1},k_{1s-\lambda_1})\left[\slashed{\Pi}_{s\perp}(k') - \slashed{\Pi}_{s\perp}(k_2)\right]\slashed{\Pi}_{s\perp}(k_1)\nonumber\\
        &+&\left.\mathcal{J}_{\,1}(k'_{s\lambda_1},k_{2s\lambda_1})\mathcal{J}_{\,2}(k_{1s\lambda_1},k_{s\lambda_1})\mathcal{J}_{\,3}(k_{2s\lambda_1},k_{1s\lambda_1})\left[-q^2_\parallel+\slashed{q}_\parallel(\slashed{\Pi}_{s\perp}(k) - \slashed{\Pi}_{s\perp}(k_1)\right]\right\} u(k,p,a).
\end{eqnarray}
Combining the denominators in Eq.~(\ref{dgamma}) and using Feynman parameters, we have (see Appendix \ref{App: Feynman} for more details)

\begin{eqnarray}
    \mathcal{V}^{\mbox{\small{ (1)}}} &=&ig^3\frac{1}{(2\pi)^2}\sum_{k_1,k_2=0}^\infty \int \frac{d^2w_\perp}{(2\pi)^2}\frac{d^2w_\parallel}{(2\pi)^2}\frac{2(2\pi)^3\delta(x+y+z-1)\delta^2(p'_\parallel-p_\parallel-l_\parallel)\delta(p'_2-p_2-l_2)}{D^3}\nonumber\\
    &\times&\sum_{\lambda_1=\pm}\bar{u}(k',p',a')\gamma^5\left\{\left[\Delta^{\lambda_1}\left(ym\slashed{\Pi}_{s\perp}(k') -y\slashed{\Pi}_{s\perp}(k')\slashed{\Pi}_{s\perp}(k)+ym\slashed{\Pi}_{s\perp}(k_2)-y\slashed{\Pi}_{s\perp}(k_2)\slashed{\Pi}_{s\perp}(k)\right.\right.\right.\nonumber\\
    &+&\left.zm\slashed{\Pi}_{s\perp}(k')+zm\slashed{\Pi}_{s\perp}(k_2)-\slashed{\Pi}_{s\perp}(k')\slashed{\Pi}_{s\perp}(k)-\slashed{\Pi}_{s\perp}(k_2)\slashed{\Pi}_{s\perp}(k)\right)\nonumber\\
    &+&\left.\Delta^{-\lambda_1}\left(z\slashed{\Pi}_{s\perp}(k')\slashed{\Pi}_{s\perp}(k')+z\slashed{\Pi}_{s\perp}(k')\slashed{\Pi}_{s\perp}(k_2)\right)\right]\mathcal{J}_{\,1}(k'_{s\lambda_1},k_{2s\lambda_1})\mathcal{J}_{\,2}(k_{1s,-\lambda_1},k_{s,-\lambda_1})\mathcal{J}_{\,3}(k_{2s,-\lambda_1},k_{1s,-\lambda_1})\nonumber\\
    &+&\Delta^{\lambda_1}\left[\slashed{\Pi}_{s\perp}(k')\slashed{\Pi}_{s\perp}(k_1) - \slashed{\Pi}_{s\perp}(k_2)\slashed{\Pi}_{s\perp}(k_1)\right]\mathcal{J}_{\,1}(k'_{s\lambda_1},k_{2s\lambda_1})\mathcal{J}_{\,2}(k_{1s\lambda_1},k_{s\lambda_1})\mathcal{J}_{\,3}(k_{2s,-\lambda_1},k_{1s,-\lambda_1})\nonumber\\
    &+&\left[\Delta^{\lambda_1}\left(-w_\parallel^2 - m^2(1-x)^2 + 2B_Q(1-x)(yk+zk')+yzm_\pi^2 -yzl_\perp^2 + ym\slashed{\Pi}_{s\perp}(k)-y\slashed{\Pi}_{s\perp}(k)\slashed{\Pi}_{s\perp}(k)\right.\right.\nonumber\\
    &-&\left.ym\slashed{\Pi}_{s\perp}(k_1)+y\slashed{\Pi}_{s\perp}(k_1)\slashed{\Pi}_{s\perp}(k)+zm\slashed{\Pi}_{s\perp}(k)-zm\slashed{\Pi}_{s\perp}(k_1)\right)\nonumber\\
    &+&\left.\left.\Delta^{-\lambda_1}\left(z\slashed{\Pi}_{s\perp}(k')\slashed{\Pi}_{s\perp}(k)-z\slashed{\Pi}_{s\perp}(k')\slashed{\Pi}_{s\perp}(k_1)\right)\right]\mathcal{J}_{\,1}(k'_{s\lambda_1},k_{2s\lambda_1})\mathcal{J}_{\,2}(k_{1s\lambda_1},k_{s\lambda_1})\mathcal{J}_{\,3}(k_{2s\lambda_1},k_{1s\lambda_1})\right\}u(k,p,a)
\end{eqnarray}

Now we perform the $w_\parallel$ integral using the well-known formulas

\begin{eqnarray}
    \int \frac{d^dw}{(2\pi)^d}\frac{1}{(w^2-\beta)^n} = \frac{(-1)^ni}{(4\pi)^\frac{d}{2}}\frac{\Gamma(n-\frac{d}{2})}{\Gamma(n)}\left(\frac{1}{\beta}\right)^{n-\frac{d}{2}}\\
    \int \frac{d^dw}{(2\pi)^d}\frac{w^2}{(w^2-\beta)^n} = \frac{(-1)^{n-1}i}{(4\pi)^\frac{d}{2}}\frac{d}{2}\frac{\Gamma(n-\frac{d}{2}-1)}{\Gamma(n)}\left(\frac{1}{\beta}\right)^{n-\frac{d}{2}-1}.
\end{eqnarray}

For our case, $\beta = -xw_\perp^2 + \eta = x\Vec{w}_\perp^2 + \eta +i\epsilon$, thus
\begin{equation}
\begin{aligned}
    \int\frac{d^2w_\parallel}{(2\pi)^2}\frac{1}{(w_\parallel^2 - \beta)^3} = \frac{-i}{8\pi}\frac{1}{\beta}\\
    \int\frac{d^2w_\parallel}{(2\pi)^2}\frac{w^2}{(w_\parallel^2 - \beta)^3} = \frac{i}{8\pi}\frac{1}{\beta}
    \label{Parallel_Integrals}
    \end{aligned}
\end{equation}
After simplifying, with $    \eta = xyp_\perp^2 + xzp'^2_\perp + yzl_\perp^2 + m^2(1-x)^2 + m_\pi^2(x-yz) + 2yk_1B_Q+2zk_2B_Q$, obtained in Appendix~\ref{App: Feynman}, Eq.~(\ref{eta_completa}), we have

\begin{eqnarray}
        \mathcal{V}^{\mbox{\small{ (1)}}} &=&ig^3\frac{(2\pi)^3\delta^2(p'_\parallel-p_\parallel-l_\parallel)\delta(p'_2-p_2-l_2)}{(2\pi)^2}\sum_{k_1,k_2=0}^\infty \int \frac{d^2w_\perp}{(2\pi)^2}\frac{2i\delta(x+y+z-1)}{8\pi}\nonumber\\
        &\times&\sum_{\lambda_1 = \pm}\frac{\bar{u}(k',p',a')\gamma^5\Delta^{\lambda_1}}{-(x\Vec{w}^2_\perp+\eta+i\epsilon)^2}\left\{\left(ym\slashed{\Pi}_{s\perp}(k') -y\slashed{\Pi}_{s\perp}(k')\slashed{\Pi}_{s\perp}(k)+ym\slashed{\Pi}_{s\perp}(k_2)-y\slashed{\Pi}_{s\perp}(k_2)\slashed{\Pi}_{s\perp}(k)\right.\right.\nonumber\\
        &+&\left.zm\slashed{\Pi}_{s\perp}(k')+zm\slashed{\Pi}_{s\perp}(k_2)-\slashed{\Pi}_{s\perp}(k')\slashed{\Pi}_{s\perp}(k)-\slashed{\Pi}_{s\perp}(k_2)\slashed{\Pi}_{s\perp}(k)\right)\mathcal{J}_{\,1}(k'_{s\lambda_1},k_{2s\lambda_1})\mathcal{J}_{\,2}(k_{1s,-\lambda_1},k_{s,-\lambda_1})\mathcal{J}_{\,3}(k_{2s,-\lambda_1},k_{1s,-\lambda_1})\nonumber\\
        &+&\left(z\slashed{\Pi}_{s\perp}(k')\slashed{\Pi}_{s\perp}(k')+z\slashed{\Pi}_{s\perp}(k')\slashed{\Pi}_{s\perp}(k_2)\right)\mathcal{J}_{\,1}(k'_{s,-\lambda_1},k_{2,-s\lambda_1})\mathcal{J}_{\,2}(k_{1s\lambda_1},k_{s\lambda_1})\mathcal{J}_{\,3}(k_{2s\lambda_1},k_{1s\lambda_1})\nonumber\\
        &+&\left(\slashed{\Pi}_{s\perp}(k')\slashed{\Pi}_{s\perp}(k_1) - \slashed{\Pi}_{s\perp}(k_2)\slashed{\Pi}_{s\perp}(k_1)\right)\mathcal{J}_{\,1}(k'_{s\lambda_1},k_{2s\lambda_1})\mathcal{J}_{\,2}(k_{1s\lambda_1},k_{s\lambda_1})\mathcal{J}_{\,3}(k_{2s,-\lambda_1},k_{1s,-\lambda_1})\nonumber\\
        &+&\left( - m^2(1-x)^2 + 2B_Q(1-x)(yk+zk')+yzm_\pi^2 -yzl_\perp^2 + ym\slashed{\Pi}_{s\perp}(k)-y\slashed{\Pi}_{s\perp}(k)\slashed{\Pi}_{s\perp}(k)\right.\nonumber\\
        &-&\left.ym\slashed{\Pi}_{s\perp}(k_1)+y\slashed{\Pi}_{s\perp}(k_1)\slashed{\Pi}_{s\perp}(k)+zm\slashed{\Pi}_{s\perp}(k)-zm\slashed{\Pi}_{s\perp}(k_1)\right)\mathcal{J}_{\,1}(k'_{s\lambda_1},k_{2s\lambda_1})\mathcal{J}_{\,2}(k_{1s\lambda_1},k_{s\lambda_1})\mathcal{J}_{\,3}(k_{2s\lambda_1},k_{1s\lambda_1})\nonumber\\
        &+&\left.\left(z\slashed{\Pi}_{s\perp}(k')\slashed{\Pi}_{s\perp}(k)-z\slashed{\Pi}_{s\perp}(k')\slashed{\Pi}_{s\perp}(k_1)\right)\mathcal{J}_{\,1}(k'_{s,-\lambda_1},k_{2s,-\lambda_1})\mathcal{J}_{\,2}(k_{1s,-\lambda_1},k_{s,-\lambda_1})\mathcal{J}_{\,3}(k_{2s,-\lambda_1},k_{1s,-\lambda_1})\right\}u(k,p,a)\nonumber\\
        &-&\bar{u}(k',p',a')\Delta^{\lambda_1}\gamma^5u(k,p,a)\frac{\mathcal{J}_{\,1}(k'_{s\lambda_1},k_{2s\lambda_1})\mathcal{J}_{\,2}(k_{1s\lambda_1},k_{s\lambda_1})\mathcal{J}_{\,3}(k_{2s\lambda_1},k_{1s\lambda_1})}{x\Vec{w}^2_\perp+\eta+i\epsilon}
    \label{final_eq}
\end{eqnarray}

Using polar coordinates for the perpendicular momenta $d^2w_{\perp}$, the angular integration can be performed, more details are shown in Appendix~\ref{App: Perp_Mom}

\begin{equation}
\begin{split}
    \int&\frac{d^2w_\perp}{(2\pi)^2}\frac{\mathcal{J}_{\,1}(k'_{s\lambda_1},k_{2s\lambda_1})\mathcal{J}_{\,2}(k_{1s\lambda_2},k_{s\lambda_2})\mathcal{J}_{\,3}(k_{2s\lambda_3},k_{1s\lambda_3})}{f(|\Vec{w}_\perp|)} =\\
    &
    \begin{dcases}
        \begin{split}
            &(2\pi)^2 e^{i\left(\frac{s}{2B_Q}\right) l_1 (p_2 + p_2')}\sqrt{\frac{k'_{s\lambda_1}!k_{1s\lambda_2}!k_{2s\lambda_3}!}{k_{2s\lambda_1}!k_{s\lambda_2}!k_{1s\lambda_3}!}}i^{k-k'}(-1)^{k-k_1}e^{-\frac{\Vec{l}_{\perp}^2}{4B_Q}}e^{-is\phi_l(k-k')}\left(\frac{\Vec{l}_\perp^2}{2B_Q}\right)^{\frac{k_1-k_2}{2}}L_{k_{2s\lambda_3}}^{k_1-k_2}\left(\frac{\Vec{l}_\perp^2}{2B_Q}\right)\\
            &\times\sum_{n=0}^{\infty}\left(\delta_{|k_2-k_1+k-k'|,2n} - sgn(k_1-k_2+k'-k)\delta_{|k_2-k_1+k-k'|,2n+1}\right)\\
            &\times\int_0^{\infty}d|\Vec{w}_\perp| \frac{|\Vec{w}_\perp|}{f(|\Vec{w}_\perp|)}e^{-\frac{\Vec{w}_{\perp}^2}{2B_Q}}\left(\frac{\Vec{w}_\perp^2}{2B_Q}\right)^{\frac{k_2-k'+k-k_1}{2}}L_{k'_{s\lambda_1}}^{k_2-k'}\left(\frac{\Vec{w}_\perp^2}{2B_Q}\right)L_{k_{1s\lambda_2}}^{k-k_1}\left(\frac{\Vec{w}_\perp^2}{2B_Q}\right)J_{|k_2-k_1+k-k'|}\left(\frac{|\Vec{w}_\perp||\Vec{l}_\perp|}{B_Q}\right)\text{ ,}
        \end{split}\qquad k \geq k_{1} \geq k_{2} \geq k'\\ 
        \begin{split}
            &(2\pi)^2 e^{i\left(\frac{s}{2B_Q}\right) l_1 (p_2 + p_2')}\sqrt{\frac{k_{2s\lambda_1}!k_{s\lambda_2}!k_{1s\lambda_3}!}{k'_{s\lambda_1}!k_{1s\lambda_2}!k_{2s\lambda_3}!}}i^{k'-k}(-1)^{k_1-k}e^{-\frac{\Vec{l}_{\perp}^2}{4B_Q}}e^{is\phi_l(k'-k)}\left(\frac{\Vec{l}_\perp^2}{2B_Q}\right)^{\frac{k_2-k_1}{2}}L_{k_{1s\lambda_3}}^{k_2-k_1}\left(\frac{\Vec{l}_\perp^2}{2B_Q}\right)\\
            &\times\sum_{n=0}^{\infty}\left(\delta_{|k_2-k_1+k-k'|,2n} - sgn(k_1-k_2+k'-k)\delta_{|k_2-k_1+k-k'|,2n+1}\right)\\
            &\times\int_0^{\infty}d|\Vec{w}_\perp| \frac{|\Vec{w}_\perp|}{f(|\Vec{w}_\perp|)}e^{-\frac{\Vec{w}_{\perp}^2}{2B_Q}}\left(\frac{\Vec{w}_\perp^2}{2B_Q}\right)^{\frac{k'-k_2+k_1-k}{2}}L_{k_{2s\lambda_1}}^{k'-k_2}\left(\frac{\Vec{w}_\perp^2}{2B_Q}\right)L_{k_{s\lambda_2}}^{k_1-k}\left(\frac{\Vec{w}_\perp^2}{2B_Q}\right)J_{|k_1-k_2+k'-k|}\left(\frac{|\Vec{w}_\perp||\Vec{l}_\perp|}{B_Q}\right)
        \end{split} \qquad k' \geq k_{2} \geq k_{1} \geq k 
    \end{dcases}
    \end{split}
    \label{Angular_integral}
\end{equation}

\end{widetext}

where
\begin{eqnarray}
  \mbox{sgn}(x) =
    \begin{cases}
    1 & \text{if } x > 0 \\
    0 & \text{if } x = 0 \\
    -1 & \text{if } x < 0
    \end{cases}
\end{eqnarray}
We notice that the one-loop vertex is written as the convolution of the three form factors coming from the vertices of the triangular diagram in Fig.~\ref{Loop}.\\\\

\section{One-loop correction in LLL using the Ritus Formalism}\label{secV}

From our result in Sec.~\ref{vertex-loop}, we take the LLL for external $u$-quarks with $k=k'=0$; then, from the restrictions in Eq.~(\ref{Angular_integral}), we get $k_1=k_2=0$. Noticing that for a $u$-quark, $s = +1$,
\begin{equation}
    k_{s\lambda} = k-\frac{1-s\lambda}{2} = 0 + \frac{\lambda-1}{2} = 
    \begin{cases}
        0 \qquad \ \ \lambda = +1\\
        -1 \qquad \lambda=-1
    \end{cases}
\end{equation}

Since $k_{+-}! = (-1)!$ is not defined, only $k_{++}$ is valid. Then the LLL correction of Eq.~(\ref{final_eq}) is given by

\begin{widetext}

\begin{equation}
    \begin{split}
        \mathcal{V}^{\mbox{\small{ (1)}}}_{LLL} =& ig^3\frac{(2\pi)^3\delta(p'^2-p^2-l^2)\delta^2(p'_\parallel-p_\parallel-l_\parallel)}{(2\pi)^2}\frac{2i}{8\pi}\int_0^1dxdydz\delta(x+y+z-1)\bar{u}(k',p',a')\Delta^{+}\gamma^5u(k,p,a)\\
        &\times\int \frac{d^2w_\perp}{(2\pi)^2}\mathcal{J}_{\,1}(k'_{++},k_{2++})\mathcal{J}_{\,2}(k_{1++},k_{++})\mathcal{J}_{\,3}(k_{2++},k_{1++})\\
        &\times\left\{\frac{-m^2(1-x)^2+yzm^2_{\pi}-yzl_{\perp}^2}{-(x\Vec{w}_\perp^2 + \eta_{LLL})^2} - \frac{1}{(x\Vec{w}_\perp^2 + \eta_{LLL})}\right\}
    \end{split}
\end{equation}

where,

\begin{equation}
    \begin{split}
        \int &\frac{d^2w_\perp}{(2\pi)^2}\mathcal{J}_{\,1}(k'_{++},k_{2++})\mathcal{J}_{\,2}(k_{1++},k_{++})\mathcal{J}_{\,3}(k_{2++},k_{1++}) \\
        &= (2\pi)^2 e^{i\frac{1}{2B_Q}l_1(p_2+p'_2)}e^{-\frac{\Vec{l_\perp^2}}{4B_Q}}\int_0^{\infty}d|\Vec{w}_\perp|\,|\Vec{w}_\perp|e^{-\frac{\Vec{w}_\perp^2}{2B_Q}}J_0\left(\frac{|\Vec{w}_\perp||\Vec{l}_\perp|}{B_Q}\right)
    \end{split}
\end{equation}

and

\begin{equation}
    \eta_{LLL} = -yz\Vec{l}_\perp^2 + m^2(1-x)^2 + m_\pi^2(x-yz)
\end{equation}

Finally after simplifying, we obtain

\begin{eqnarray}
        \mathcal{V}^{\mbox{\small{ (1)}}}_{LLL} &=& g^3(2\pi)^3\delta(p'^2-p^2-l^2)\delta^2(p'_\parallel-p_\parallel-l_\parallel)\bar{u}(0,p',a')\Delta^{+}\gamma^5u(0,p,a)e^{i\frac{1}{2B_Q}l^1(p^2+p'^2)}\nonumber\\
        &\times& \frac{1}{4\pi}e^{-\frac{\Vec{l_\perp^2}}{4B_Q}}\int_0^1dxdydz\int_0^{\infty}d|\Vec{w}_\perp|\,\delta(x+y+z-1)|\Vec{w}_\perp|e^{-\frac{\Vec{w}_\perp^2}{2B_Q}}J_0\left(\frac{|\Vec{w}_\perp||\Vec{l}_\perp|}{B_Q}\right)\left\{\frac{x(m^2_{\pi}+\Vec{w}^2_{\perp})}{(x\Vec{w}_\perp^2 + \eta_{LLL} +i\epsilon)^2}\right\}
    \label{LLL_Result_1}
\end{eqnarray}
\end{widetext}

\begin{widetext}

\section{One-loop correction in the LLL using Schwinger's proper-time formalism}\label{Schwing}

To cross-check our previous result using  an independent calculation, here we calculate the one-loop correction to the $u\pi_{0}u$ vertex using the Schwinger proper time method in the LLL. We start again with the one-loop correction for the vertex given by Eq.~(\ref{vertex}). We start by using the expression for the LLL case, namely
\begin{equation}
    \mathcal{V}^{\mbox{\small{ (1)}}}_{LLL} = i^3\int d^4xd^4yd^4z \bar{U}_{LLL}(p',x)(-ig\gamma^5)S^{LLL}_f(x,z)(-ig\gamma^5)S^{LLL}_f(z,y)(-ig\gamma^5)D(y,x)U_{LLL}(y,p)e^{-il\cdot z}.
    \label{vertex_LLL}
\end{equation}
In coordinate space, the LLL Schwinger propagator has the form
\begin{equation}
    S^{LLL}_{f}(x,y)=e^{i\Phi(x,y)}\overline{S^{LLL}_{f}}(x-y)
    \label{Schwinger_Propagator_New}
\end{equation}
where, in LG2 the Schwinger phase is given by,
\begin{equation}
    \Phi(x,y)=\frac{B_{Q}}{2}(x^{1}+y^{1})(x^2-y^2)\,,
    \label{Schwinger_Phase_New}
\end{equation}
and the translationally invariant part is
\begin{eqnarray}
    \overline{S^{L\!L\!L}_{f}}(x-y)\!&=&\!\frac{B_{Q}}{2\pi}e^{-\frac{B_{Q}}{4}(\vec{x}-\vec{y})^{2}_{\perp}}
    \!\!\int\!\!\frac{d^{2}q_{\parallel}}{(2\pi)^2}e^{-iq_{\parallel}\cdot(x-y)_{\parallel}}\!\frac{i(\slashed{q}_{\!\parallel}\!+\!m_{u})}{q^{2}_{\!\parallel}\!-m^{2}_{u}\!+\!i\epsilon}\Delta^{+}
    \label{Translationally_Invariant_Part_New}
\end{eqnarray}
The explicit form for the propagator shown in Eq.~(\ref{Schwinger_Propagator_New}) is derived in Appendix \ref{App: Schwinger Phase Propagator}. Substituting Eq.~(\ref{Schwinger_Propagator_New}) into Eq.~(\ref{vertex_LLL}), taking into account Eqs.~(\ref{Schwinger_Phase_New}) and (\ref{Translationally_Invariant_Part_New}) and integrating over all space variables we obtain
\begin{equation}
\begin{split}
    \mathcal{V}^{\mbox{\small{ (1)}}}_{LLL}=&-ig^{3}\int d^{2}r_{\parallel}\int d^{2}r^{\prime}_{\parallel}\int d^{4}\ell\frac{1}{(\ell^{2}-m^{2}_{\pi}+i\epsilon)(r^{2}_{\parallel}-m^{2}_{u}+i\epsilon)(r^{\prime2}_{\parallel}-m^{2}_{u}+i\epsilon)}e^{-\frac{\vec{\ell}_{\perp}^{2}}{2B_{Q}}}e^{\frac{iq^{1}(p^{2}+p^{\prime2})}{2B_{Q}}}\\
    &\times e^{-\frac{\vec{q}_{\perp}^{2}}{4B_{Q}}}e^{\frac{iq^{1}\ell^{2}}{B_{Q}}}e^{\frac{-iq^{2}\ell^{1}}{B_{Q}}}\delta(p^{\prime2}-p^{2}-q^{2})\delta(p^{0}+\ell^{0}-r^{\prime0})\delta(p^{\prime0}+\ell^{0}-r^{0})\delta(p^{3}+\ell^{3}-r^{\prime3})\delta(p^{\prime3}+\ell^{3}-r^{3})\\
    &\times\delta(r^{0}-r^{\prime0}-q^{0})\delta(r^{3}-r^{\prime3}-q^{3})\bar{u}(0,p^{\prime3},a^{\prime})\Delta^{+}\gamma^{5}(\slashed{r}_{\parallel}+m_{u})\Delta^{+}\gamma^{5}(\slashed{r}^{\prime}_{\parallel}+m_{u})\Delta^{+}\gamma^{5}\Delta^{+}u(0,p^{3},a)\,,
    \label{Vertex_Spatially_Integrated}
    \end{split}
\end{equation}
Performing the integrations over $r_{\parallel}$ and $r'_{\parallel}$ in Eq.~(\ref{Vertex_Spatially_Integrated}) we get the following constraints
\begin{equation}
\begin{aligned}
r_{\parallel} &= p^{\prime}_{\parallel}+\ell_{\parallel}\,\\
r^{\prime}_{\parallel} &= p_{\parallel}+\ell_{\parallel}\,.
\end{aligned}
\end{equation}
Thus, Eq.~(\ref{Vertex_Spatially_Integrated}) simplifies to
\begin{equation}
\begin{split}
    \mathcal{V}^{\mbox{\small{ (1)}}}_{LLL}=-&ig^{3}\delta(p^{\prime}_{\parallel}-p_{\parallel}-q_{\parallel})\delta(p^{\prime2}-p^{2}-q^{2})e^{-\frac{\vec{q}_{\perp}^{2}}{4B_{Q}}}e^{\frac{iq^{1}(p^{2}+p^{\prime2})}{2B_{Q}}}\int d^{4}\ell\frac{1}{(\ell^{2}-m^{2}_{\pi}+i\epsilon)(r^{2}_{\parallel}-m^{2}_{u}+i\epsilon)(r^{\prime2}_{\parallel}-m^{2}_{u}+i\epsilon)}\\
    &\times e^{-\frac{\vec{\ell}_{\perp}^{2}}{2B_{Q}}}e^{\frac{iq^{1}\ell^{2}}{B_{Q}}}e^{\frac{-iq^{2}\ell^{1}}{B_{Q}}}\bar{u}(0,p^{\prime3},a^{\prime})\Delta^{+}\gamma^{5}(\slashed{p}^{\prime}_{\parallel}+\slashed{l}_{\parallel}+m_{u})\Delta^{+}\gamma^{5}(\slashed{p}_{\parallel}+\slashed{l}_{\parallel}+m_{u})\Delta^{+}\gamma^{5}\Delta^{+}u(0,p^{3},a)\,.
    \end{split}
    \label{Vertex_Spatially_r_rpime_Integrated}
\end{equation}
We now combine the denominators in Eq.~(\ref{Vertex_Spatially_r_rpime_Integrated}) using Feynman parameters obtaining the following result (see also Eq.~(\ref{Feynman_Parameters}) in Appendix B)
\begin{equation}
    \frac{1}{(\ell^{2}-m^{2}_{\pi}+i\epsilon)((p_{\parallel}+\ell_{\parallel})^{2}-m^{2}_{u}+i\epsilon)((p^{\prime}_{\parallel}+\ell_{\parallel})^{2}-m^{2}_{u}+i\epsilon)}=\int_{0}^{1}dxdydz\frac{2}{D^{3}}\delta(x+y+z-1)\,,
    \label{Denominator}
\end{equation}
where,
\begin{equation}
\begin{aligned}
D &= w^{2}_{\parallel}+xw^{2}_{\perp}-\eta_{LLL}+i\epsilon\,,\\
w_{\parallel} &= \ell_{\parallel}+(yp_{\parallel}+zp^{\prime}_{\parallel})\,,\\
w_{\perp} &= \ell_{\perp}\\
\eta_{LLL} &= -yz\vec{q}^{2}_{\perp}+m^{2}_{u}(1-x)^{2}+m^{2}_{\pi}(x-yz)\,.
\end{aligned}
\end{equation}
Note that with these new variables, the measure in Eq.~(\ref{Vertex_Spatially_r_rpime_Integrated}) remains invariant, that is
\begin{equation}
    d^{4}\ell\equiv d^{2}\ell_{\parallel}d^{2}\ell_{\perp}=d^{2}w_{\parallel}d^{2}w_{\perp}\,.
    \label{Measure}
\end{equation}
We will now rewrite the numerator in Eq.~(\ref{Vertex_Spatially_r_rpime_Integrated}) using the new variable $w_{\parallel}$ and Feynman parameters. Before doing so, let us simplify the spinorial structure:
\begin{equation}
    \bar{u}\Delta^{+}\gamma^{5}(\slashed{p}^{\prime}_{\parallel}+\slashed{l}_{\parallel}+m_{u})\Delta^{+}\gamma^{5}(\slashed{p}_{\parallel}+\slashed{l}_{\parallel}+m_{u})\Delta^{+}\gamma^{5}\Delta^{+}u=-\bar{u}\Delta^{+}\gamma^{5}(\slashed{p}^{\prime}_{\parallel}+\slashed{l}_{\parallel}+m_{u})(\slashed{p}_{\parallel}+\slashed{l}_{\parallel}-m_{u})u\,,
    \label{Numerator_1}
\end{equation}
substituting the value $\ell_{\parallel}=w_{\parallel}-yp_{\parallel}-zp^{\prime}_{\parallel}$ in Eq.~(\ref{Numerator_1}) and simplifying using the Dirac equation, its adjoint, and the on-shell conditions for external particles, we have that the numerator which we denote by $N$ reduces to
\begin{equation}
    N=w^{2}_{\parallel}-(x+xy+xz)m^{2}_{u}-yzm^{2}_{\pi}+m^{2}_{u}\,.
    \label{Numerator_2}
\end{equation}
Inserting Eqs.~(\ref{Denominator}), (\ref{Measure}), (\ref{Numerator_1}) and (\ref{Numerator_2}) in Eq.~(\ref{Vertex_Spatially_r_rpime_Integrated}) we get
\begin{equation}
\begin{split}
    \mathcal{V}^{\mbox{\small{ (1)}}}_{LLL}=&2ig^{3}\delta(p^{\prime}_{\parallel}-p_{\parallel}-q_{\parallel})\delta(p^{\prime2}-p^{2}-q^{2})e^{-\frac{\vec{q}_{\perp}^{2}}{4B_{Q}}}e^{\frac{iq^{1}(p^{2}+p^{\prime2})}{2B_{Q}}}\int_{0}^{1}dxdydz\delta(x+y+z-1)\bar{u}\Delta^{+}\gamma^{5}u\\
    &\times\int d^{2}w_{\perp}\times e^{-\frac{\vec{w}_{\perp}^{2}}{2B_{Q}}}e^{\frac{iq^{1}w^{2}}{B_{Q}}}e^{\frac{-iq^{2}w^{1}}{B_{Q}}}\int d^{2}w_{\parallel}\frac{w^{2}_{\parallel}-(x+xy+xz)m^{2}_{u}-yzm^{2}_{\pi}+m^{2}_{u}}{(w^{2}_{\parallel}+xw^{2}_{\perp}-\eta_{LLL}+i\epsilon)^{3}}\,.
    \label{Vertex_Simplified}
\end{split}    
\end{equation}
Next, we evaluate the parallel integral in Eq.~(\ref{Vertex_Simplified}) using the formulas in Eq.~(\ref{Parallel_Integrals}), obtaining the result
\begin{equation}
\begin{split}
    \mathcal{V}^{\mbox{\small{ (1)}}}_{LLL}=&-\frac{1}{4\pi}g^{3}(2\pi)^{2}\delta(p^{\prime}_{\parallel}-p_{\parallel}-q_{\parallel})\delta(p^{\prime2}-p^{2}-q^{2})e^{-\frac{\vec{q}_{\perp}^{2}}{4B_{Q}}}e^{\frac{iq^{1}(p^{2}+p^{\prime2})}{2B_{Q}}}\int_{0}^{1}dxdydz\delta(x+y+z-1)\bar{u}\Delta^{+}\gamma^{5}u\\
    &\times\int d^{2}w_{\perp} e^{-\frac{\vec{w}_{\perp}^{2}}{2B_{Q}}}e^{\frac{iq^{1}w^{2}}{B_{Q}}}e^{\frac{-iq^{2}w^{1}}{B_{Q}}}\left(\frac{1}{\eta_{LLL}+x\vec{w}^{2}_{\perp}}+\frac{(x+xy+xz)m^{2}_{u}+yzm^{2}_{\pi}-m^{2}_{u}}{(\eta_{LLL}+x\vec{w}^{2}_{\perp})^{2}}\right)\,.
    \label{Vertex_Simplified_2}
\end{split}    
\end{equation}
Finally, we compute the perpendicular integral in Eq.~(\ref{Vertex_Simplified_2}) using polar coordinates, yielding
\begin{equation}
\begin{split}
    \mathcal{V}^{\mbox{\small{ (1)}}}_{LLL}=&-\frac{1}{4\pi}g^{3}(2\pi)^{3}\delta(p^{\prime}_{\parallel}-p_{\parallel}-q_{\parallel})\delta(p^{\prime2}-p^{2}-q^{2})e^{-\frac{\vec{q}_{\perp}^{2}}{4B_{Q}}}e^{\frac{iq^{1}(p^{2}+p^{\prime2})}{2B_{Q}}}\int_{0}^{1}dxdydz\delta(x+y+z-1)\bar{u}\Delta^{+}\gamma^{5}u\\
    &\times\int d|\vec{w}_{\perp}||\vec{w}_{\perp}| e^{-\frac{|\vec{w}_{\perp}^{2}|}{2B_{Q}}}J_{0}\left(\frac{|\vec{w}_{\perp}||\vec{q}_{\perp}|}{B_{Q}}\right)\left(\frac{1}{\eta_{LLL}+x\vec{w}^{2}_{\perp}}+\frac{(x+xy+xz)m^{2}_{u}+yzm^{2}_{\pi}-m^{2}_{u}}{(\eta_{LLL}+x\vec{w}^{2}_{\perp})^{2}}\right)\,.
\end{split}    
\label{LLL_Result_2}
\end{equation}
Eq.~(\ref{LLL_Result_2}) is identical to Eq.~(\ref{LLL_Result_1}), confirming that both methods yield the same result.
\end{widetext}

\begin{figure}[t!]
    \centering
    \includegraphics[width=1\linewidth]{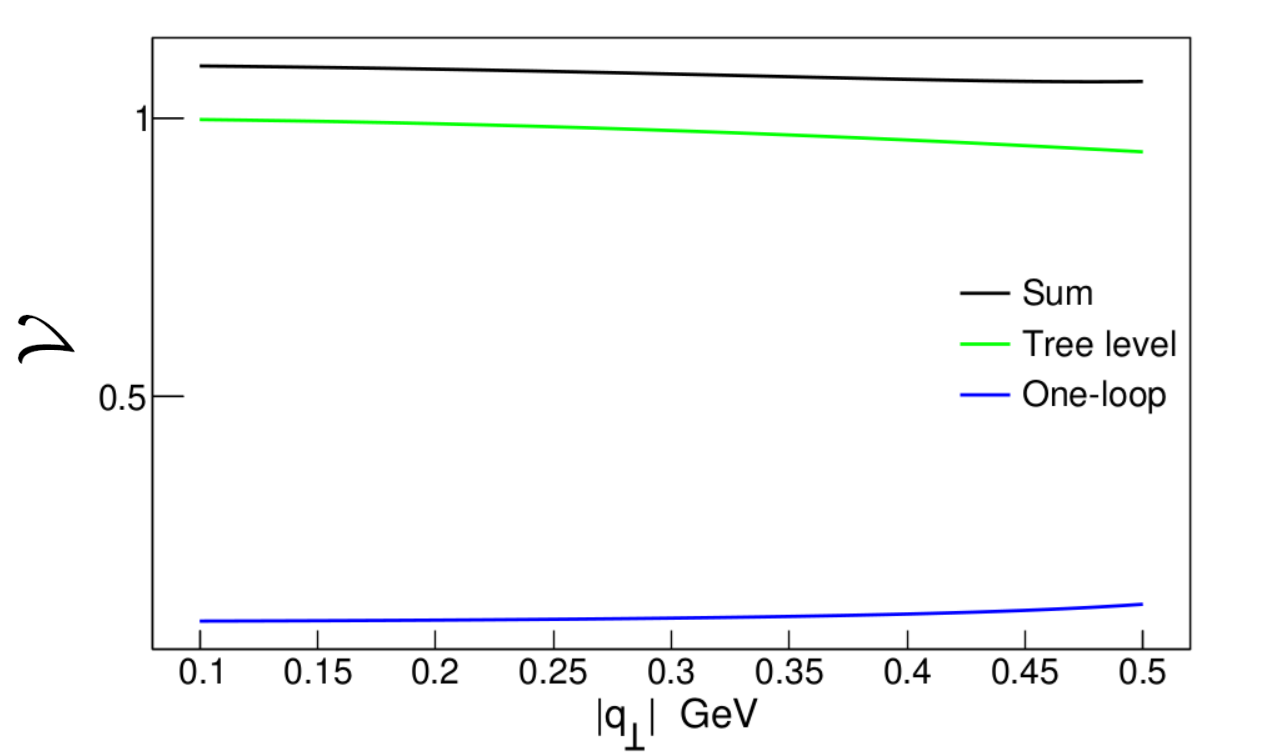}
    \caption{Strength of the form factor for the $u\pi_0u$ coupling as a function of the quark transverse momentum at tree-level (green curve), one-loop order (blue curve) and their sum (black curve), where the overall momentum-conserving delta function and the associated $(2\pi)^4$ normalization factor have been factored out. The form factors are computed for the case where the quark occupies the LLL, and the constants used for numerical calculation are $B_Q=1$ GeV$^2$, $m_f=0.3$ GeV and $m_\pi=0.14$ GeV.}
    \label{Results_1}
\end{figure}

\section{Results}\label{results}

\begin{figure}[t!]
    \centering
    \includegraphics[width=1\linewidth]{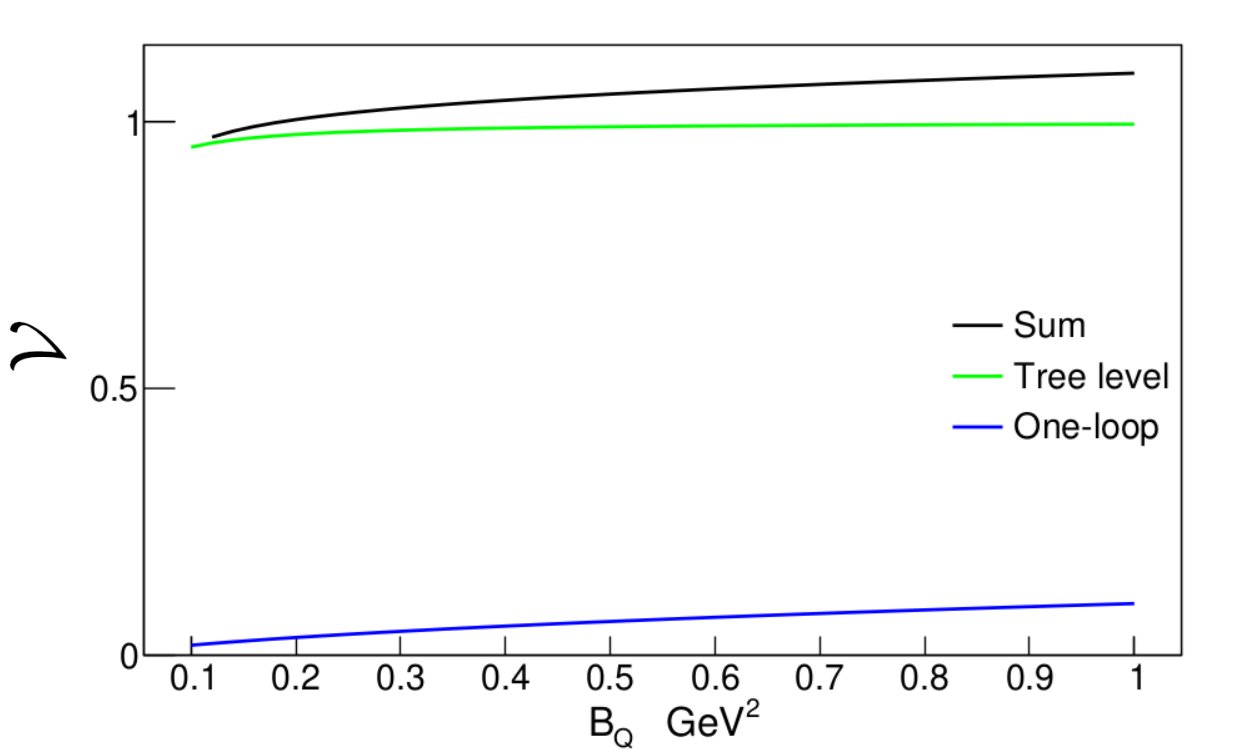}
    \caption{Strength of the form factor for the $u\pi_0u$ coupling as a function of the magnetic field intensity at tree-level (green curve), one-loop order (blue curve) and their sum (black curve), where the overall momentum-conserving delta function and the associated $(2\pi)^4$ normalization factor have been factored out. The form factors are computed for the case where the quark occupies the LLL, and the constants used for numerical calculation are $|q_\perp|=m_\pi$, $m_f=0.3$ GeV and $m_\pi=0.14$ GeV. }
    \label{Results_2}
\end{figure}

Before presenting the results for the form factors, a few words are in order. First, notice that in vacuum, the $u\pi_0 u$ vertex is given by $-ig\gamma^5$; therefore, the \lq\lq form factor" is just the momentum independent value \lq\lq 1". Also, recall that in vacuum, the description of the asymptotic states is given in terms of plane waves. In contrast, in the magnetic case, we cannot use plane waves for the quark asymptotic states, instead, we are required to use the Ritus states whose quantum numbers are $\breve{p}=(k,p_2,p_3)$ and $\breve{p}'=(k',p'_2,p'_3)$. In the vacuum plane-wave picture, one labels these states in terms of the transverse momenta $p_\perp$ and $p'_\perp$ instead of the Landau levels $k$ and $k'$, respectively. The relationship between the quantum numbers in one case and the other is given by
\begin{eqnarray}
p_\perp^2&=&2kB_Q\nonumber\\
p_\perp'^2&=&2k'B_Q.
\label{relation}
\end{eqnarray}
Therefore, taking $B_Q\to 0$ at fixed $k$, $k'$, does not correspond to taking physical $p_\perp$, $p'_\perp$ values in the vacuum theory, instead, this corresponds to sending $p_\perp, p'_\perp\to 0$. Should we wanted to reproduce the usual plane-wave amplitude at given fixed $p_\perp$, $p'_\perp$, we would need to take $B_Q\to 0$, $k,k'\to\infty$ keeping $2kB_Q=p_\perp^2$ and $2k'B_Q=p_\perp'^2$, fixed. Physically, this means that the only realizable way to match the Ritus description with the plane-wave description corresponds to the situation where the effect of the magnetic field is negligible, namely for very large values of the Landau level indices. The immediate implication is that the external-state basis with a fixed Landau level, especially the LLL, does not smoothly match onto the plane-wave vacuum basis as $B_Q\to 0$, so the magnetic form factors, defined as the functions multiplying $-ig\gamma_5$, do not go to 1 as $B_Q\to 0$. To get an adequate picture, one would need to sum over all Landau levels to expect the correct behavior as $B_Q\to 0$.

How can one then define a sensible magnetic field-modified form factor? Recall that in vacuum the $\pi_0q\bar{q}$ form factor is defined by the plane-wave $S$-matrix element
\begin{eqnarray}
\langle u(p)\bar{u}(p')|\pi_0(q)\rangle=igF(B_Q=0;q)\bar{u}(p')\gamma_5u(p).
\end{eqnarray}

By definition, $F(B_Q=0;q)=1$. Within a background magnetic field, the closest definition of an $S$-matrix element calls for the computation of the magnetic field modified vertex between a quark and an antiquark with given Ritus-quantum numbers, sandwiched between plane wave states, and, as argued above, performing the sum over all Landau levels.

Let us look at the implications of this definition of the form factors. Let
\begin{eqnarray}
    \Big\{\big|\bar{p}=(p_0,k,p_2,p_3)\bigr>_{B_Q}\Big\}, 
\end{eqnarray}
be a complete set of Ritus eigenstates at fixed $B_Q$. Completeness implies that
\begin{eqnarray}
\sumint_{\bar{p}}\big|\bar{p}\bigr>_{B_Q\ B_Q}\! \bigl<\bar{p}\big| =1
\end{eqnarray}

Now, consider the plane-wave matrix element of the magnetic field dependent vertex operator $\big<p'\big|{\mathcal{\hat{V}}}\big|p\big>$. Inserting the identity in the Ritus basis, we get
\begin{eqnarray}
\big<p'\big|{\mathcal{\hat{V}}}\big|p\big>=\sumint_{\bar{p}'}\sumint_{\bar{p}}\big<p'\big|\bar{p}'\bigr>_{B_Q\ B_Q}\! \bigl<\bar{p}'\big|{\mathcal{\hat{V}}}|\bar{p}\bigr>_{B_Q\ B_Q}\! \bigl<\bar{p}\big|p\big>.\nonumber\\
\label{matrixelementexact}
\end{eqnarray}

Notice that Eq.~(\ref{matrixelementexact}) is exact. First, consider this expression computed at tree-level, namely ${\mathcal{\hat{V}}}={\mathcal{\hat{V}}^{\mbox{\small{ (tree)}}}}=ig\gamma_5$. Notice that this operator is local and $B_Q$-independent. In this case, it can be factored out, resulting in the full sum over Landau levels just reconstructing the plane-wave resolution of the identity, namely,
\begin{eqnarray}
\big<p'\big|{\mathcal{\hat{V}}^{\mbox{\small{ (tree)}}}}\big|p\big>=\big<p'\big|{\mathcal{\hat{V}}^{\mbox{\small{ (tree)}}}}\big|p\big>_{\mbox{\small{vacuum}}}.
\end{eqnarray}
As a consequence 
\begin{eqnarray}
    F^{\mbox{\small{(tree)}}}(B_Q;q)=1,
\end{eqnarray}
exactly, for all $B_Q$ and all momenta. Therefore, there is no magnetic modification at tree-level in this fully Landau-level summed computation of the form factor defined between plane-wave states. At one-loop, the situation changes. Since charged particles propagate in the loop, their propagators depend on $B_Q$ and thus, even after summing over Landau levels, the loop integrals retain a genuine $B_Q$-dependence. Therefore, to compute the modification of the form factor, it is sensible to sum the corresponding contributions, one Landau level at a time.

We now present our numerical results for the magnetic field-driven corrections to the $u\pi_{0}u$ vertex . We show the tree and one-loop corrections for the case where the quark occupies the LLL. Figure~\ref{Results_1} shows the tree-level $\mathcal{V^{\mbox{\small{(tree)}}}}$
and the one-loop $\mathcal{V}^{\mbox{\small{(1)}}}$ form factors, as well as their sum, as functions of the pion momentum $q_\perp$. As can be observed, the tree-level form factor slightly decreases while the one-loop form factor slightly increases, resulting in their sum slightly decreasing with increasing $q_\perp$. This behavior is expected since, for larger pion momentum, the magnetic field, which dominates the dynamics when $q_\perp$ is small, becomes comparatively less influential. The tree-level form factor is always smaller than the one-loop form factor. Notice also that the one-loop correction is almost an order of magnitude larger than the tree-level one. For the calculation, we used $B_Q = 1$ GeV$^2$, $m_{f} = 0.3$ GeV and $m_\pi = 0.14$ GeV. Figure~\ref{Results_2} shows the tree-level $\mathcal{V^{\mbox{\small{(tree)}}}}$, and the one-loop $\mathcal{V}^{\mbox{\small{(1)}}}$ form factors, as well as their sum, as functions of the magnetic field $B_Q$. Both corrections slightly increase with the field strength. Once again, we notice that the one-loop correction is almost an order of magnitude larger than the tree-level one. For the calculation, we used $|q_\perp | = m_\pi$, $m_{f} = 0.3$ GeV and $m_\pi = 0.14$ GeV. 

\section{Summary and conclusions}\label{conclusions}

In this work, we have studied the magnetic-field-induced corrections to the $u\pi^{0}u$ vertex using the LSMq, both at tree-level and at one-loop order. The magnetic field effects are introduced both in terms of the Ritus functions, which describe the charged states in the presence of the external field, and in the case of the one-loop calculation, also using the magnetic field dependent quark propagators. Choosing the quark to occupy the LLL, we have shown that the general calculation in the Ritus formalism and the one using the Schwinger proper time formalism coincide. We have found that the sum of the tree-level and one-loop form factors slightly decreases with increasing $q_\perp$. This illustrates the fact that at larger pion momentum, the magnetic field, which dominates the dynamics when $q_\perp$ is small, becomes comparatively less influential. On the other hand, the form factor increases with increasing field strength. We also observe that, for the chosen input parameters, the one-loop correction in the LLL approximation is almost an order of magnitude larger than the tree-level correction. The results of this work can be used to study the selection rules for transitions involving quarks and neutral pions for different physical processes of interest. This is work in progress that will be reported elsewhere.

\section*{Acknowledgments}

A.A. wishes to thank the colleagues and staff of Universidade de São Paulo and of Instituto de F\'isica Te\'orica, UNESP for their kind hospitality during a sabbatical stay. A.A. also acknowledges support from the PASPA program of the Direcci\'on General de Asuntos del Personal Acad\'emico (DGAPA) of the Universidad Nacional Aut\'onoma de M\'exico (UNAM) for the sabbatical stay during which this research was carried out. Support for this work has been received in part by a DGAPA-PAPIIT-UNAM grant number IG100826 and from Secretar\'\i a de Ciencia, Humanidades, Tecnolog\'\i a e Innovaci\'on (SECIHTI) M\'exico grant numbers CIORGANISMOS-2025-17, CBF-2025-G-1718 and CBF-2026-465. This study was financed in part by the São Paulo Research Foundation (FAPESP), Brasil, Process Number 2024/18493-8 and Process Number 2026/06478-0. J.R. acknowledges support from the program estancias posdoctorales por M\'exico and from the SNII program, both granted by SECIHTI.


\begin{widetext}

\appendix

\section{Spatial Integrals} \label{App: Spatial_Integrals}

The general form of the three spatial integrals of Sec.~\ref{vertex-loop} is

\begin{equation}
    I = \int d^4x\,\bar{\E}(x,\bar{v})\E(x,\bar{w})e^{\alpha i hx}
\end{equation}

where v,w,h are momenta and $\alpha=\pm$

\begin{equation}
    I = \sum_{\lambda=\pm}\Delta^{\lambda}\int d^4x \F^*_Q(x,\bar{v}_{\lambda})\F_Q(x,\bar{w}_{\lambda})e^{\alpha i hx}
\end{equation}

\begin{equation}
    I = \sum_{\lambda=\pm}\Delta^{\lambda}(2\pi)^3\delta^2(v_\parallel-w_\parallel+\alpha h_\parallel)\delta(v_2-w_2+\alpha h_2)\,J_{k_{s\lambda},k'_{s\lambda}}(v_\perp,w_\perp,h_\perp)
\end{equation}

Defining

\begin{equation}
    J_{k,k'}(v_\perp,w_\perp,h_\perp) = N_{k_{s\lambda}}N_{k'_{s\lambda}}\int d^4x D_{k_{s\lambda}}\left[\sqrt{2B_Q x_1} - s\sqrt{\frac{2}{B_Q}}v_2\right]D_{k'_{s\lambda}}\left[\sqrt{2B_Q x_1} - s\sqrt{\frac{2}{B_Q}}w_2\right] e^{-\alpha i h_1x_1}
\end{equation}

Using the formula
\begin{equation}
    \int_{-\infty}^{\infty}d\psi\, e^{i\gamma \psi}D_{l}(\eta-\psi)D_n
(\eta + \psi) = 
\begin{cases}
    (-1)^l \sqrt{2\pi}\,l!e^{-\frac{\gamma^2 + \eta^2}{2}}(i\gamma + \eta)^{n-l}L^{n-l}_{l}(\eta^2 + \gamma^2)\quad n\geq l \\
    (-1)^n \sqrt{2\pi}\,n!e^{-\frac{\gamma^2 + \eta^2}{2}}(-i\gamma + \eta)^{l-n}L^{l-n}_{n}(\eta^2 + \gamma^2)\quad l\geq n
\end{cases}
\end{equation}

And the property, $D_{l}(\psi-\eta) = (-1)^lD_{l}(\eta-\psi) $. For our case

\begin{eqnarray}
    \psi = \sqrt{2B_Q x_1} - \frac{s}{\sqrt{2B_Q}}(v_2+w_2)\\
    \eta = \frac{s}{\sqrt{2B_Q}}(v_2 - w_2) = \frac{-s\alpha h_2}{\sqrt{2B_Q}}\\
    \gamma = -\frac{\alpha h_1}{\sqrt{2B_Q}}\\
    \frac{\eta^2 + \gamma^2}{2} = \frac{h_2^2 + h_1^2}{4B_Q}\\
    \eta + i \gamma = \frac{-s\alpha h_2 - i\alpha h_1}{\sqrt{2B_Q}}
\end{eqnarray}

Then

\begin{equation}
    J_{k,k'}(v_\perp,w_\perp,h_\perp) = e^{-\frac{i\alpha s h_1(v_2 + w_2)}{2B_Q}}G_{k.k'}(h_\perp),
\end{equation}

where

\begin{equation}
    G_{k.k'}(h_\perp) = 2\pi e^{-\frac{h_1^2+h_2^2}{4B_Q}} 
    \begin{cases}
        \sqrt{\frac{k!}{k'!}}\left(\frac{-s\alpha h_2 - i \alpha h_1}{\sqrt{2B_Q}}\right)^{k'-k} L_k^{k'-k}\left(\frac{h_1^2 + h_2^2}{2B_Q}  \right) \quad k' \geq k\\
        \sqrt{\frac{k'!}{k!}}\left(\frac{s\alpha h_2 - i \alpha h_1}{\sqrt{2B_Q}}\right)^{k-k'} L_{k'}^{k-k'}\left(\frac{h_1^2 + h_2^2}{2B_Q}  \right) \quad k \geq k'
    \end{cases}
\end{equation}

Defining $\Vec{h}_\perp = (h_1,h_2)$ and $tan(\phi_h) = \frac{h_2}{h_1}$ then $e^{\pm is\phi_h} = \frac{1}{|\Vec{h}_\perp|}(h_1 \pm i s h_2)$

\begin{equation}
    G_{k.k'}(h_\perp) = 2\pi e^{-\frac{\Vec{h}^2_\perp}{4B_Q}} 
    \begin{cases}
        \sqrt{\frac{k!}{k'!}}(-i\alpha)^{k'-k}e^{-is\phi_h(k'-k)}\left(\frac{\Vec{h}^2_\perp}{2B_Q}\right)^{\frac{k'-k}{2}} L_k^{k'-k}\left(\frac{h^2_\perp}{2B_Q}  \right) \quad k' \geq k\\
        \sqrt{\frac{k'!}{k!}}(-i\alpha)^{k-k'}e^{is\phi_h(k-k')}\left(\frac{\Vec{h}^2_\perp}{2B_Q}\right)^{\frac{k-k'}{2}} L_{k'}^{k-k'}\left(\frac{\Vec{h}^2_\perp}{2B_Q}  \right) \quad k \geq k'
    \end{cases}
\end{equation}

Finally,

\begin{equation}
    I = \sum_{\lambda=\pm}\Delta^{\lambda}(2\pi)^3\delta^2(v_\parallel-w_\parallel+\alpha h_\parallel)\delta(v_2-w_2+\alpha h_2)\,J_{k_{s\lambda},k'_{s\lambda}}(v_\perp,w_\perp,h_\perp)
\end{equation}

\begin{equation}
    J_{k,k'}(v_\perp,w_\perp,h_\perp) = e^{-\frac{i\alpha s h_1(v_2 + w_2)}{2B_Q}}G_{k.k'}(h_\perp)
\end{equation}

\begin{equation}
    G_{k.k'}(h_\perp) = 2\pi e^{-\frac{\Vec{h}^2_\perp}{4B_Q}} 
    \begin{cases}
        \sqrt{\frac{k!}{k'!}}(-i\alpha)^{k'-k}e^{-is\phi_h(k'-k)}\left(\frac{\Vec{h}^2_\perp}{2B_Q}\right)^{\frac{k'-k}{2}} L_k^{k'-k}\left(\frac{h^2_\perp}{2B_Q}  \right) \quad k' \geq k\\
        \sqrt{\frac{k'!}{k!}}(-i\alpha)^{k-k'}e^{is\phi_h(k-k')}\left(\frac{\Vec{h}^2_\perp}{2B_Q}\right)^{\frac{k-k'}{2}} L_{k'}^{k-k'}\left(\frac{\Vec{h}^2_\perp}{2B_Q}  \right) \quad k \geq k'
    \end{cases}
\end{equation}

\section{Feynman parameters}\label{App: Feynman}

Here we perform the algebra involving the Feynman parameters for the calculation of Sec.~\ref{vertex-loop}. To combine the three denominators, we use
\begin{equation}
    \begin{aligned}
        \frac{1}{A_1 A_2 A_3} &=  \int_0^1 dx_1dx_2dx_3 \delta(x_1 + x_2 + x_3 - 1) \frac{2}{(x_1A_1 + x_2A_2 + x_3A_3)^3} \\ &= \int_0^1 dx_1dx_2dx_3 \delta(x_1 + x_2 + x_3 - 1) \frac{2}{D^3}\,,
    \end{aligned}
    \label{Feynman_Parameters}
\end{equation}
where,

\begin{equation}
D = x(q^2 - m_\pi) + y[(p+q)_{\parallel}^2 - m^2 -2k_1 B_Q] + z[(p'+q)_{\parallel}^2 - m^2 -2k_2 B_Q] + (x+y+z)i\epsilon\,,
\end{equation}
simplifying the previous expression, we have

\begin{equation}
D = q^{2}_{\parallel} + xq_{\perp}^2 + 2(yp_\parallel +zp'_{\parallel})\cdot q_\parallel  + yp_{\parallel}^2 + z {p'}^2_{\parallel} -ym^2-zm^2-xm_\pi^2-2yk_1B_Q-2zk_2B_Q + i\epsilon\,.
\end{equation}

It is convenient to define the following variables
\begin{equation}
    \begin{cases}
        w_\parallel = q_\parallel + (yp_\parallel + zp'_\parallel)\\
        w_\perp = q_\perp
    \end{cases}
\end{equation}
Using the previous definitions, the denominator $D$ takes the form
\begin{equation}
    D = xw_\perp^2 + w_\parallel^2 - \eta + i\epsilon
\end{equation}
where 
\begin{equation}\label{eta_completa}
    \eta = xyp_\perp^2 + xzp'^2_\perp + yzl_\perp^2 + m^2(1-x)^2 + m_\pi^2(x-yz) + 2yk_1B_Q+2zk_2B_Q
\end{equation}

It is convenient to make the following change of variable in the integral of Eq.~(\ref{dgamma})
\begin{gather}
    q_\parallel = w_\parallel - (yp_\parallel +zp'_\parallel)\,,
\end{gather}

therefore Eq.~(\ref{dgamma}) takes the form
\begin{equation}
    \begin{split}
        \mathcal{V}^{\mbox{\small{ (1)}}} =& ig^3 \frac{1}{(2\pi)^2}\sum_{k_1,k_2 = 0}^{\infty}\int \frac{d^4q}{(2\pi)^4} \frac{(2\pi)^3 \delta^2(p'_\parallel-p_\parallel-l_\parallel)\delta(p_2-p'_2-l_2)\sum_{\lambda_1=\pm}\bar{u}(k',p',a')\gamma^5\Delta^{\lambda_1}}{((p+q)_{\parallel}^2 -m^2 -2k_1B_Q +i\epsilon)((p'+q)_{\parallel}^2 -m^2 -2k_2B_Q +i\epsilon)(q^2-m_\pi^2 +i\epsilon)}\\ 
        &\left\{\mathcal{J}_{1}(k'_{s\lambda_1},k_{2s\lambda_1})\mathcal{J}_{2}(k_{1s,-\lambda_1},k_{s,-\lambda_1})\mathcal{J}_{3}(k_{2s,-\lambda_1},k_{1s,-\lambda_1})\left.\left[(\slashed{\Pi}_{s\perp}(k') + \slashed{\Pi}_{s\perp}(k_2))(\slashed{w}_\parallel - (y\slashed{p}_\parallel +z\slashed{p}'_\parallel))\right.\right.\right.\\        
        & \left.-(\slashed{\Pi}_{s\perp}(k')+\slashed{\Pi}_{s\perp}(k_2))\slashed{\Pi}_{s\perp}(k)\right]- \mathcal{J}_{1}(k'_{s\lambda_1},k_{2s\lambda_1})\mathcal{J}_{2}(k_{1s\lambda_1},k_{s\lambda_1})\mathcal{J}_{3}(k_{2s-\lambda_1},k_{1s-\lambda_1})(\slashed{\Pi}_{s\perp}(k') - \slashed{\Pi}_{s\perp}(k_2))\slashed{\Pi}_{s\perp}(k_1)\\
        &\left.\mathcal{J}_{1}(k'_{s\lambda_1},k_{2s\lambda_1})\mathcal{J}_{2}(k_{1s\lambda_1},k_{s\lambda_1})\mathcal{J}_{3}(k_{2s\lambda_1},k_{1s\lambda_1})\left[-(w_\parallel - (yp_\parallel +zp'_\parallel))^2+(\slashed{w}_\parallel - (y\slashed{p}_\parallel +z\slashed{p}'_\parallel))(\slashed{\Pi}_{s\perp}(k) - \slashed{\Pi}_{s\perp}(k_1)\right]\right\} u(k,p,a)
    \end{split}\label{dgamma2}
\end{equation}


\section{Perpendicular momenta integration}\label{App: Perp_Mom}

For the perpendicular momenta integration, there are two distinct denominators, both depending only on the radial component of $\Vec{w}_\perp$. They are denoted as $f(|\Vec{w}_\perp|)$

\begin{equation}
    \begin{split}
        I&=\int\frac{d^2w_\perp}{(2\pi)^2}\frac{\mathcal{J}_{1}(k'_{s\lambda_1},k_{2s\lambda_1})\mathcal{J}_{2}(k_{1s\lambda_2},k_{s\lambda_2})\mathcal{J}_{3}(k_{2s\lambda_3},k_{1s\lambda_3})}{f(|\Vec{w}_\perp|)} \\
        &= \int\frac{d^2w_\perp}{(2\pi)^2} e^{-is\left(\frac{w_1(2p'_2+w_2)}{2B_Q}\right)}e^{is\left(\frac{w_1(2p_2+w_2)}{2B_Q}\right)}e^{is\left(\frac{l_1(p_2 + p'_2+2w_2)}{2B_Q}\right)}\frac{G_{k'_{s\lambda_1},k_{2s\lambda_1}}(w_\perp)G_{k_{1s\lambda_2},k_{s\lambda_2}}(w_\perp)G_{k_{2s\lambda_3},k_{1s\lambda_3}}(l_\perp)}{f(|\Vec{w}_\perp|)}
    \end{split}
\end{equation}

Simplifying the exponetials

\begin{equation}
    e^{-i s \frac{w_1[2p_2' + w_2]}{2B_Q}} \, e^{i s \frac{w_1[2p_2 + w_2]}{2B_Q}} \, e^{i s \frac{l_1[p_2 + p_2' + 2w_2]}{2B_Q}} 
\end{equation}

Using $\delta(p_2' - p_2 - l_2)$ we get

\begin{equation}
    e^{-i\frac{s}{B_Q}(w_1 l_2 - w_2 l_1)} \, e^{i\left(\frac{s}{2B_Q}\right) l_1 (p_2 + p_2')}
\end{equation}

Defining

\begin{equation}
    \begin{cases}
w_1 = |\vec{w}_\perp|\cos\phi_w \\
w_2 = |\vec{w}_\perp|\sin\phi_w
\end{cases}
\qquad
\begin{cases}
l_1 = |\vec{l}_\perp|\cos\phi_l \\
l_2 = |\vec{l}_\perp|\sin\phi_l
\end{cases}
\end{equation}

After simplifying we get

\begin{equation}
    e^{i \frac{s|\vec{w}_\perp||\vec{l}_\perp|}{B_Q}\sin(\phi_w - \phi_l)} \, e^{i\left(\frac{s}{2B_Q}\right) l_1 (p_2 + p_2')}
\end{equation}

Then simplifying the product of the G functions,

\begin{equation}
    \begin{split}
        \mathcal{G} &= G_{k'_{s\lambda_1},k_{2s\lambda_1}}(w_\perp)G_{k_{1s\lambda_2},k_{s\lambda_2}}(w_\perp)G_{k_{2s\lambda_3},k_{1s\lambda_3}}(l_\perp)\\
        &=(2\pi)e^{-\frac{\Vec{w}_{\perp}^2}{4B_Q}}\begin{cases}
        \sqrt{\frac{k'_{s\lambda_1}!}{k_{2s\lambda_1}!}}i^{k_{2s\lambda_1}-k'_{s\lambda_1}}e^{-is\phi_w(k_{2s\lambda_1}-k'_{s\lambda_1})}\left(\frac{\Vec{w}^2_\perp}{2B_Q}\right)^{\frac{k_{2s\lambda_1}-k'_{s\lambda_1}}{2}} L_{k'_{s\lambda_1}}^{k_{2s\lambda_1}-k'_{s\lambda_1}}\left(\frac{w^2_\perp}{2B_Q}  \right) \quad k_{2s\lambda_1} \geq k'_{s\lambda_1}\\
        \sqrt{\frac{k_{2s\lambda_1}!}{k'_{s\lambda_1}!}}i^{k'_{s\lambda_1} - k_{2s\lambda_1}}e^{is\phi_w(k'_{s\lambda_1} - k_{2s\lambda_1})}\left(\frac{\Vec{w}^2_\perp}{2B_Q}\right)^{\frac{k'_{s\lambda_1} - k_{2s\lambda_1}}{2}} L_{k_{2s\lambda_1}}^{k'_{s\lambda_1} - k_{2s\lambda_1}}\left(\frac{\Vec{w}^2_\perp}{2B_Q}  \right) \quad k'_{s\lambda_1} \geq k_{2s\lambda_1}
    \end{cases}\\
    &\times(2\pi)e^{-\frac{\Vec{w}_{\perp}^2}{4B_Q}}\begin{cases}
        \sqrt{\frac{k_{1s\lambda_2}!}{k_{s\lambda_2}!}}(-i)^{k_{s\lambda_2} - k_{1s\lambda_2}}e^{-is\phi_w(k_{s\lambda_2} - k_{1s\lambda_2})}\left(\frac{\Vec{w}^2_\perp}{2B_Q}\right)^{\frac{k_{s\lambda_2} - k_{1s\lambda_2}}{2}} L_{k_{1s\lambda_2}}^{k_{s\lambda_2} - k_{1s\lambda_2}}\left(\frac{w^2_\perp}{2B_Q}  \right) \quad k_{s\lambda_2}  \geq k_{1s\lambda_2}\\
        \sqrt{\frac{k_{s\lambda_2}!}{k_{1s\lambda_2}!}}(-i)^{k_{1s\lambda_2} - k_{s\lambda_2}}e^{is\phi_w(k_{1s\lambda_2} - k_{s\lambda_2})}\left(\frac{\Vec{w}^2_\perp}{2B_Q}\right)^{\frac{k_{1s\lambda_2} - k_{s\lambda_2}}{2}} L_{k_{s\lambda_2}}^{k_{1s\lambda_2} - k_{s\lambda_2}}\left(\frac{\Vec{w}^2_\perp}{2B_Q}  \right) \quad k_{1s\lambda_2} \geq k_{s\lambda_2}
    \end{cases}\\
    &\times(2\pi)e^{-\frac{\Vec{l}_{\perp}^2}{4B_Q}}\begin{cases}
        \sqrt{\frac{k_{2s\lambda_2}!}{k_{1s\lambda_2}!}}i^{k_{1s\lambda_3} - k_{2s\lambda_3}}e^{-is\phi_l(k_{1s\lambda_3} - k_{2s\lambda_3})}\left(\frac{\Vec{l}^2_\perp}{2B_Q}\right)^{\frac{k_{1s\lambda_3} - k_{2s\lambda_3}}{2}} L_{k_{2s\lambda_3}}^{k_{1s\lambda_3} - k_{2s\lambda_3}}\left(\frac{l^2_\perp}{2B_Q}  \right) \quad k_{1s\lambda_3} \geq  k_{2s\lambda_3}\\
        \sqrt{\frac{k_{1s\lambda_3}!}{k_{2s\lambda_3}!}}i^{k_{2s\lambda_3} - k_{1s\lambda_3}}e^{is\phi_l(k_{2s\lambda_3} - k_{1s\lambda_3})}\left(\frac{\Vec{l}^2_\perp}{2B_Q}\right)^{\frac{k_{2s\lambda_3} - k_{1s\lambda_3}}{2}} L_{k_{1s\lambda_3}}^{k_{2s\lambda_3} - k_{1s\lambda_3}}\left(\frac{\Vec{l}^2_\perp}{2B_Q}  \right) \quad k_{2s\lambda_3} \geq k_{1s\lambda_3}
    \end{cases}
    \end{split}
\end{equation}

Notice that $k_{s\lambda} - k'_{s\lambda} = k-\frac{1-s\lambda}{2} - (k'-\frac{1-s\lambda}{2}) = k-k'$, and $k_{s\lambda} \geq k'_{s\lambda} \rightarrow \, k \geq k'$. There are only two consistent possibilities:

\begin{equation}
    \begin{split}
        &\begin{cases}
        k_{2s\lambda_1} \geq k'_{s\lambda_1}\\
        k_{s\lambda_2} \geq k_{1s\lambda_2}\\
        k_{1s\lambda_3} \geq k_{2s\lambda_3}
        \end{cases}
        \Longrightarrow
        k \geq k_{1} \geq k_{2} \geq k'\\
        \\
         &\begin{cases}
        k'_{s\lambda_1} \geq k_{2s\lambda_1}\\
        k_{1s\lambda_2} \geq k_{s\lambda_2}\\
        k_{2s\lambda_3} \geq k_{1s\lambda_3}
        \end{cases}
        \Longrightarrow
        k' \geq k_{2} \geq k_{1} \geq k\\
    \end{split}
\end{equation}
\subsection{Case \texorpdfstring{$k \geq k_{1} \geq k_{2} \geq k'$}
{Case k >= k1 >= k2 >= k'}}
\begin{equation}
    \begin{split}
        \mathcal{G} =& (2\pi)^3\sqrt{\frac{k'_{s\lambda_1}!k_{1s\lambda_2}!k_{2s\lambda_3}!}{k_{2s\lambda_1}!k_{s\lambda_2}!k_{1s\lambda_3}!}}i^{k-k'}(-1)^{k-k_1}e^{-\frac{\Vec{l}_{\perp}^2}{4B_Q}}e^{-is\phi_l(k_1-k_2)}\left(\frac{\Vec{l}_\perp^2}{2B_Q}\right)^{\frac{k_1-k_2}{2}}L_{k_{2s\lambda_3}}^{k_1-k_2}\left(\frac{\Vec{l}_\perp^2}{2B_Q}\right)e^{-is\phi_w(k_2-k'+k-k_1)}\\
        &e^{-\frac{\Vec{w}_{\perp}^2}{2B_Q}}\left(\frac{\Vec{w}_\perp^2}{2B_Q}\right)^{\frac{k_2-k'+k-k_1}{2}}L_{k'_{s\lambda_1}}^{k_2-k'}\left(\frac{\Vec{w}_\perp^2}{2B_Q}\right)L_{k_{1s\lambda_2}}^{k-k_1}\left(\frac{\Vec{w}_\perp^2}{2B_Q}\right)
    \end{split}
\end{equation}

Making a change of variable to polar coordinates and the angular change $\phi'_w = \phi_w - \phi_l$, $d^2w_\perp = |\Vec{w}_\perp|d|\Vec{w}_\perp|d\phi'_w$, we get

\begin{equation}
    \begin{split}
        I =& (2\pi)e^{i\left(\frac{s}{2B_Q}\right) l_1 (p_2 + p_2')}\sqrt{\frac{k'_{s\lambda_1}!k_{1s\lambda_2}!k_{2s\lambda_3}!}{k_{2s\lambda_1}!k_{s\lambda_2}!k_{1s\lambda_3}!}}i^{k-k'}(-1)^{k-k_1}e^{-\frac{\Vec{l}_{\perp}^2}{4B_Q}}e^{-is\phi_l(k-k')}\left(\frac{\Vec{l}_\perp^2}{2B_Q}\right)^{\frac{k_1-k_2}{2}}L_{k_{2s\lambda_3}}^{k_1-k_2}\left(\frac{\Vec{l}_\perp^2}{2B_Q}\right)\\
        &\times\int_0^{\infty}d|\Vec{w}_\perp| \int_0^{2\pi} d\phi'_w \frac{|\Vec{w}_\perp|}{f(|\Vec{w}_\perp|)}e^{i \frac{s|\vec{w}_\perp||\vec{l}_\perp|}{B_Q}\sin(\phi'_w)} e^{-is\phi'_w(k_2-k'+k-k_1)}\\
        &\times e^{-\frac{\Vec{w}_{\perp}^2}{2B_Q}}\left(\frac{\Vec{w}_\perp^2}{2B_Q}\right)^{\frac{k_2-k'+k-k_1}{2}}L_{k'_{s\lambda_1}}^{k_2-k'}\left(\frac{\Vec{w}_\perp^2}{2B_Q}\right)L_{k_{1s\lambda_2}}^{k-k_1}\left(\frac{\Vec{w}_\perp^2}{2B_Q}\right)
    \end{split}
\end{equation}

Using the following identities, $\alpha=\pm$

\begin{equation}
    e^{\alpha i z\sin(\phi)} = J_0(z) + 2\sum_{n=1}^{\infty}J_{2n}(z)\cos(2n\phi) + \alpha2i\sum_{n=0}^{\infty}J_{2n+1}(z)\sin((2n+1)\phi)
\end{equation}

\begin{equation}
    \delta_{k,j} = \frac{1}{2\pi}\int_{0}^{2\pi}d\phi \, e^{i(k-j)\phi}
\end{equation}

We get our new identity

\begin{equation}
    \int_{0}^{2\pi}d\phi\, e^{im\phi+\alpha iz\sin{\phi}} = (2\pi)\, \sum_{n=0}^{\infty} J_{|m|}(z)\left(\delta_{|m|,2n} - \alpha\, sgn(m)\delta_{|m|,2n+1}\right)
\end{equation}

Then, for our case we have

\begin{equation}
    \begin{split}
        \int_{0}^{2\pi}d\phi'\,&e^{i \frac{s|\vec{w}_\perp||\vec{l}_\perp|}{B_Q}\sin(\phi'_w)} e^{-is\phi'_w(k_2-k'+k-k_1)}\\
        &= 2\pi\, \sum_{n=0}^{\infty}J_{|k_2-k_1+k-k'|}\left(\frac{|\Vec{w}_\perp||\Vec{l}_\perp|}{B_Q}\right)\left(\delta_{|k_2-k_1+k-k'|,2n} - sgn(k_1-k_2+k'-k)\delta_{|k_2-k_1+k-k'|,2n+1}\right)
    \end{split}
\end{equation}

Finally, we obtain

\begin{equation}
    \begin{split}
        &\int\frac{d^2w_\perp}{(2\pi)^2}\frac{\mathcal{J}_{1}(k'_{s\lambda_1},k_{2s\lambda_1})\mathcal{J}_{2}(k_{1s\lambda_2},k_{s\lambda_2})\mathcal{J}_{3}(k_{2s\lambda_3},k_{1s\lambda_3})}{f(|\Vec{w}_\perp|)}\\
        &=(2\pi)^2 e^{i\left(\frac{s}{2B_Q}\right) l_1 (p_2 + p_2')}\sqrt{\frac{k'_{s\lambda_1}!k_{1s\lambda_2}!k_{2s\lambda_3}!}{k_{2s\lambda_1}!k_{s\lambda_2}!k_{1s\lambda_3}!}}i^{k-k'}(-1)^{k-k_1}e^{-\frac{\Vec{l}_{\perp}^2}{4B_Q}}e^{-is\phi_l(k-k')}\left(\frac{\Vec{l}_\perp^2}{2B_Q}\right)^{\frac{k_1-k_2}{2}}L_{k_{2s\lambda_3}}^{k_1-k_2}\left(\frac{\Vec{l}_\perp^2}{2B_Q}\right)\\
        &\times\sum_{n=0}^{\infty}\left(\delta_{|k_2-k_1+k-k'|,2n} - sgn(k_1-k_2+k'-k)\delta_{|k_2-k_1+k-k'|,2n+1}\right)\\
        &\times\int_0^{\infty}d|\Vec{w}_\perp| \frac{|\Vec{w}_\perp|}{f(|\Vec{w}_\perp|)}e^{-\frac{\Vec{w}_{\perp}^2}{2B_Q}}\left(\frac{\Vec{w}_\perp^2}{2B_Q}\right)^{\frac{k_2-k'+k-k_1}{2}}L_{k'_{s\lambda_1}}^{k_2-k'}\left(\frac{\Vec{w}_\perp^2}{2B_Q}\right)L_{k_{1s\lambda_2}}^{k-k_1}\left(\frac{\Vec{w}_\perp^2}{2B_Q}\right)J_{|k_2-k_1+k-k'|}\left(\frac{|\Vec{w}_\perp||\Vec{l}_\perp|}{B_Q}\right)
    \end{split}
\end{equation}

\subsection{\texorpdfstring
{Case $k' \geq k_{2} \geq k_{1} \geq k$}
{Case k' >= k2 >= k1 >= k}}
Performing the same procedure as before

\begin{equation}
    \begin{split}
        &\int\frac{d^2w_\perp}{(2\pi)^2}\frac{\mathcal{J}_{1}(k'_{s\lambda_1},k_{2s\lambda_1})\mathcal{J}_{2}(k_{1s\lambda_2},k_{s\lambda_2})\mathcal{J}_{3}(k_{2s\lambda_3},k_{1s\lambda_3})}{f(|\Vec{w}_\perp|)}\\
        &=(2\pi)^2 e^{i\left(\frac{s}{2B_Q}\right) l_1 (p_2 + p_2')}\sqrt{\frac{k_{2s\lambda_1}!k_{s\lambda_2}!k_{1s\lambda_3}!}{k'_{s\lambda_1}!k_{1s\lambda_2}!k_{2s\lambda_3}!}}i^{k'-k}(-1)^{k_1-k}e^{-\frac{\Vec{l}_{\perp}^2}{4B_Q}}e^{is\phi_l(k'-k)}\left(\frac{\Vec{l}_\perp^2}{2B_Q}\right)^{\frac{k_2-k_1}{2}}L_{k_{1s\lambda_3}}^{k_2-k_1}\left(\frac{\Vec{l}_\perp^2}{2B_Q}\right)\\
        &\times\sum_{n=0}^{\infty}\left(\delta_{|k_2-k_1+k-k'|,2n} - sgn(k_1-k_2+k'-k)\delta_{|k_2-k_1+k-k'|,2n+1}\right)\\
        &\times\int_0^{\infty}d|\Vec{w}_\perp| \frac{|\Vec{w}_\perp|}{f(|\Vec{w}_\perp|)}e^{-\frac{\Vec{w}_{\perp}^2}{2B_Q}}\left(\frac{\Vec{w}_\perp^2}{2B_Q}\right)^{\frac{k'-k_2+k_1-k}{2}}L_{k_{2s\lambda_1}}^{k'-k_2}\left(\frac{\Vec{w}_\perp^2}{2B_Q}\right)L_{k_{s\lambda_2}}^{k_1-k}\left(\frac{\Vec{w}_\perp^2}{2B_Q}\right)J_{|k_1-k_2+k'-k|}\left(\frac{|\Vec{w}_\perp||\Vec{l}_\perp|}{B_Q}\right)
    \end{split}
\end{equation}

\section{Schwinger propagator in the LLL approximation}\label{App: Schwinger Phase Propagator}
Here we derive the explicit form for the Schwinger propagator in the LLL approximation that is shown in Eqs.~(\ref{Schwinger_Propagator_New}),~(\ref{Schwinger_Phase_New}) and~(\ref{Translationally_Invariant_Part_New}). We start with the general expression for the fermion propagator in Ritus space, which is given by Eq.~(\ref{fermion_propagator}) and we also write it here for convenience
\begin{equation}
    S_f(x,y) = \sumint_{\bar{q}}\E^{Q}(x,\bar{q})S_f(k,q_{\parallel})\bar{\E}^{Q}(y,\bar{q})\,,
    \label{Fermion_Propagator_Complete}
\end{equation}
where
\begin{equation}
S_f(k,q_{\parallel})=\frac{\slashed{\Pi}_s + m_u}{q_{\parallel}^2 - m_u^2-2kB_Q+i\epsilon}\,,
\label{Fermion_Propagator}
\end{equation}
and
\begin{equation}
    \Pi_s^\mu(q_0,k,q_3) = (q_0,0,-s\sqrt{2kB_Q},q_3)\,.
    \label{Momentum_Ritus}
\end{equation}
The Ritus functions for fermions are given by
\begin{equation}
    \E^{Q}(x,\bar{q}) = \sum_{\lambda = \pm} \Delta^\lambda\F_{Q}(x,\bar{q}_\lambda)\,, 
    \label{Ritus_Lepton_Function}
\end{equation}
with the conjugated function
\begin{equation}
    {\bar{\E}^{Q}}(y,\bar{q})=\gamma^{0}{\E}^{Q}(y,\bar{q})^{\dagger}\gamma^{0}\,,
    \label{Conjugated_Ritus_Lepton_Function}
\end{equation}
and the spin projectors
\begin{equation}
\begin{aligned}
\Delta^{\lambda} &=\frac{(1 + i\lambda\gamma^1 \gamma^2)}{2}\,.
\end{aligned}
\end{equation}
We work in the LG2, where we have
\begin{equation}
      \sumint_{\bar{q}} \equiv \frac{1}{2\pi}\sum_{k = 0}^{\infty}\int \frac{dq_0}{2\pi}\frac{dq_2}{2\pi}\frac{dq_3}{2\pi}\,,
      \label{Measure_LG2}
\end{equation}
and
\begin{equation}
    \overline{q}_\lambda = (q_0,k_{s\lambda},q_2,q_3)\,,
\end{equation}
with
\begin{equation}
    k_{s\lambda} = k - \frac{(1-s\lambda)}{2}\,.
    \label{kS_Lambda}
\end{equation}
The explicit form of the Ritus functions in this gauge was shown in Eq.~(\ref{Ritus_Function}), and we reproduce it here for ease of reference
\begin{eqnarray}
     \F_{Q}(x,\bar{q}) = N_k e^{-i(q_0x_0 - q_2x_2-x_3x_3)}D_k\left(\sqrt{2B_Q}\left(x_1 - \frac{sq_2}{B_Q}\right)\right)\,,
\end{eqnarray}
where
\begin{eqnarray}
N_k &=& (4\pi B_Q)^{1/4}/\sqrt{k!}\,,\nonumber\\
D_k(x) &=& 2^{-k/2}e^{-x^2/4}H_k(x/\sqrt{2})\,.
\end{eqnarray}
Expanding Eq.~(\ref{Ritus_Lepton_Function}), we get
\begin{equation}
\begin{split}
    \E^{Q}(x,\bar{q}) =& \Delta^{+}\F_{Q}(x,\bar{q}_{+})+\Delta^{-}\F_{Q}(x,\bar{q}_{-})\\
    =&\Delta^{+}\F_{Q}(x,q_{0},k_{++},q_{2},q_{3})+\Delta^{-}\F_{Q}(x,q_{0},k_{s-},q_{2},q_{3})\,. 
    \end{split}
    \label{Ritus_Spninor}
\end{equation}
Since we are considering the $u$ quark, we have $s=1$. Thus, in the LLL approximation, Eq.~(\ref{kS_Lambda}) reduces to
\begin{equation}
    k_{s\lambda}  = 0 + \frac{\lambda-1}{2} = 
    \begin{cases}
        k_{++}=0\,, \\
        k_{+-}-1\,.
    \end{cases}
\end{equation}
Moreover
\begin{equation}
   \F_{Q}(x,q_{0},0,q_{2},q_{3})=(4\pi B_{Q})^{\frac{1}{4}}e^{-i(q_{0}x_{0}-q_{2}x_{2}-q_{3}x_{3})}e^{-\frac{B_{Q}}{2}\left(x_{1}-\frac{q_{2}}{B_{Q}}\right)^{2}}\,,
   \label{Ritus_LLL1}
\end{equation}
and
\begin{equation}
    \F_{Q}(x,q_{0},-1,q_{2},q_{3})=0\,.
    \label{Ritus_LLL2}
\end{equation}
In the previous two equations, we have used the fact that $H_{0}(x)=1$ and $H_{-1}(x)=0$. Substituting Eqs.~(\ref{Ritus_LLL1}) and (\ref{Ritus_LLL2}) in Eq.~(\ref{Ritus_Spninor}), we obtain
\begin{equation}
    \E^{Q}(x,\bar{q})=(4\pi B_{Q})^{\frac{1}{4}}e^{-i(q_{0}x_{0}-q_{2}x_{2}-q_{3}x_{3})}e^{-\frac{B_{Q}}{2}\left(x_{1}-\frac{q_{2}}{B_{Q}}\right)^{2}}\Delta^{+}\,.
    \label{Ritus_lepton_LLL_1}
\end{equation}
Similarly, for Eq.~(\ref{Conjugated_Ritus_Lepton_Function}), we get
\begin{equation}
    {\bar{\E}^{Q}}(y,\bar{q})=(4\pi B_{Q})^{\frac{1}{4}}e^{+i(q_{0}y_{0}-q_{2}y_{2}-q_{3}y_{3})}e^{-\frac{B_{Q}}{2}\left(y_{1}-\frac{q_{2}}{B_{Q}}\right)^{2}}\Delta^{+}\,.
    \label{Ritus_Lepton_LLL_2}
\end{equation}
In the LLL approximation, Eq.~(\ref{Fermion_Propagator}) takes the form
\begin{equation}
S_f(0,q_{\parallel})=\frac{\slashed{q} + m_u}{q_{\parallel}^2 - m_u^2+i\epsilon}\,,
\label{Fermion_Propagator_LLL}
\end{equation}
Similarly, Eq.~(\ref{Momentum_Ritus}) becomes
\begin{equation}
     \Pi_s^\mu(q_0,0,q_3) = (q_0,0,0,q_3)=q_{\parallel}\,.
     \label{Momentum_Ritus_LLL}
\end{equation}
In this regime we also have that Eq.~(\ref{Measure_LG2}) simplifies to
\begin{equation}
     \sumint_{\bar{q}}=\frac{1}{2\pi}\int \frac{dq_{0}}{2\pi}\frac{dq_{2}}{2\pi}\frac{dq_{3}}{2\pi}\,.
     \label{Measure_LLL}
\end{equation}
Substituting Eqs.~(\ref{Ritus_lepton_LLL_1}), (\ref{Ritus_Lepton_LLL_2}), (\ref{Fermion_Propagator_LLL}), (\ref{Momentum_Ritus_LLL}) and (\ref{Measure_LLL}) into Eq.~(\ref{Fermion_Propagator_Complete}), we find that the fermion propagator in the LLL approximation reduces to
\begin{equation}
     S^{LLL}_f(x,y)=\frac{(4\pi B_{Q})^{1/2}}{2\pi}\int \frac{dq_{0}}{2\pi}\frac{dq_{2}}{2\pi}\frac{dq_{3}}{2\pi}e^{i[q_{0}(y_{0}-x_{0})-q_{2}(y_{2}-x_{2})-q_{3}(y_{3}-x_{3})]}e^{-\frac{B_{Q}}{2}\left(x_{1}-\frac{q_{2}}{B_{Q}}\right)^{2}}e^{-\frac{B_{Q}}{2}\left(y_{1}-\frac{q_{2}}{B_{Q}}\right)^{2}}\frac{\slashed{q} + m_u}{q_{\parallel}^2 - m_u^2+i\epsilon}\Delta^{+}\,.
\end{equation}
Finally, integrating the previous equation with respect to $q^{2}$, we obtain
\begin{eqnarray}
    S^{L\!L\!L}_{f}(x,y)\!&=&\!\frac{B_{Q}}{2\pi}e^{\frac{iB_{Q}}{2}(x_{1}+y_{1})(x_2-y_2)}e^{-\frac{B_{Q}}{4}(\vec{x}-\vec{y})^{2}_{\perp}}
    \!\!\int\!\!\frac{d^{2}q_{\parallel}}{(2\pi)^2}e^{-iq_{\parallel}\cdot(x-y)_{\parallel}}\!\frac{i(\slashed{q}_{\!\parallel}\!+\!m_{u})}{q^{2}_{\!\parallel}\!-m^{2}_{u}\!+\!i\epsilon}\Delta^{+}\,,
\end{eqnarray}
which can be written compactly as
\begin{equation}
    S^{LLL}_{f}(x,y)=e^{i\Phi(x,y)}\overline{S^{LLL}_{f}}(x-y)\,,
    \label{Schwinger_Propagator}
\end{equation}
where the Schwinger phase is given by
\begin{equation}
    \Phi(x,y)=\frac{B_{Q}}{2}(x_{1}+y_{1})(x_2-y_2)\,,
    \label{Schwinger_Phase}
\end{equation}
and the translationally invariant part is
\begin{eqnarray}
    \overline{S^{L\!L\!L}_{f}}(x-y)\!&=&\!\frac{B_{Q}}{2\pi}e^{-\frac{B_{Q}}{4}(\vec{x}-\vec{y})^{2}_{\perp}}
    \!\!\int\!\!\frac{d^{2}q_{\parallel}}{(2\pi)^2}e^{-iq_{\parallel}\cdot(x-y)_{\parallel}}\!\frac{i(\slashed{q}_{\!\parallel}\!+\!m_{u})}{q^{2}_{\!\parallel}\!-m^{2}_{u}\!+\!i\epsilon}\Delta^{+}\,.
    \label{Translationally_Invariant_Part}
\end{eqnarray}
\end{widetext}

\bibliography{biblio}{}

@article{Schwinger:1948iu,
    author = "Schwinger, Julian S.",
    title = "{On Quantum electrodynamics and the magnetic moment of the electron}",
    doi = "10.1103/PhysRev.73.416",
    journal = "Phys. Rev.",
    volume = "73",
    pages = "416--417",
    year = "1948"
}

@article{Kharzeev:2007jp,
    author = "Kharzeev, Dmitri E. and McLerran, Larry D. and Warringa, Harmen J.",
    title = "{The Effects of topological charge change in heavy ion collisions: 'Event by event P and CP violation'}",
    eprint = "0711.0950",
    archivePrefix = "arXiv",
    primaryClass = "hep-ph",
    doi = "10.1016/j.nuclphysa.2008.02.298",
    journal = "Nucl. Phys. A",
    volume = "803",
    pages = "227--253",
    year = "2008"
}

@article{Skokov:2009qp,
    author = "Skokov, V. and Illarionov, A. Yu. and Toneev, V.",
    title = "{Estimate of the magnetic field strength in heavy-ion collisions}",
    eprint = "0907.1396",
    archivePrefix = "arXiv",
    primaryClass = "nucl-th",
    doi = "10.1142/S0217751X09047570",
    journal = "Int. J. Mod. Phys. A",
    volume = "24",
    pages = "5925--5932",
    year = "2009"
}

@article{Voronyuk:2011jd,
    author = "Voronyuk, V. and Toneev, V. D. and Cassing, W. and Bratkovskaya, E. L. and Konchakovski, V. P. and Voloshin, S. A.",
    title = "{(Electro-)Magnetic field evolution in relativistic heavy-ion collisions}",
    eprint = "1103.4239",
    archivePrefix = "arXiv",
    primaryClass = "nucl-th",
    doi = "10.1103/PhysRevC.83.054911",
    journal = "Phys. Rev. C",
    volume = "83",
    pages = "054911",
    year = "2011"
}

@article{Duncan:1992hi,
    author = "Duncan, Robert C. and Thompson, Christopher",
    title = "{Formation of very strongly magnetized neutron stars - implications for gamma-ray bursts}",
    doi = "10.1086/186413",
    journal = "Astrophys. J. Lett.",
    volume = "392",
    pages = "L9",
    year = "1992"
}

@article{Kouveliotou:1998ze,
    author = "Kouveliotou, C. and others",
    title = "{An X-ray pulsar with a superstrong magnetic field in the soft gamma-ray repeater SGR 1806-20.}",
    doi = "10.1038/30410",
    journal = "Nature",
    volume = "393",
    pages = "235--237",
    year = "1998"
}

@article{Grasso:2000wj,
    author = "Grasso, Dario and Rubinstein, Hector R.",
    title = "{Magnetic fields in the early universe}",
    eprint = "astro-ph/0009061",
    archivePrefix = "arXiv",
    reportNumber = "DFPD-00-TH-35",
    doi = "10.1016/S0370-1573(00)00110-1",
    journal = "Phys. Rept.",
    volume = "348",
    pages = "163--266",
    year = "2001"
}

@article{Gusynin:1999pq,
    author = "Gusynin, V. P. and Miransky, V. A. and Shovkovy, I. A.",
    title = "{Theory of the magnetic catalysis of chiral symmetry breaking in QED}",
    eprint = "hep-ph/9908320",
    archivePrefix = "arXiv",
    reportNumber = "UCTP-103-99",
    doi = "10.1016/S0550-3213(99)00573-8",
    journal = "Nucl. Phys. B",
    volume = "563",
    pages = "361--389",
    year = "1999"
}

@article{Ayala:2006sv,
    author = "Ayala, Alejandro and Bashir, Adnan and Raya, Alfredo and Rojas, Eduardo",
    title = "{Dynamical mass generation in strongly coupled quantum electrodynamics with weak magnetic fields}",
    eprint = "hep-ph/0602209",
    archivePrefix = "arXiv",
    doi = "10.1103/PhysRevD.73.105009",
    journal = "Phys. Rev. D",
    volume = "73",
    pages = "105009",
    year = "2006"
}

@article{Bali:2011qj,
    author = "Bali, G. S. and Bruckmann, F. and Endrodi, G. and Fodor, Z. and Katz, S. D. and Krieg, S. and Schafer, A. and Szabo, K. K.",
    title = "{The QCD phase diagram for external magnetic fields}",
    eprint = "1111.4956",
    archivePrefix = "arXiv",
    primaryClass = "hep-lat",
    doi = "10.1007/JHEP02(2012)044",
    journal = "JHEP",
    volume = "02",
    pages = "044",
    year = "2012"
}

@article{Bali:2012zg,
    author = "Bali, G. S. and Bruckmann, F. and Endrodi, G. and Fodor, Z. and Katz, S. D. and Schafer, A.",
    title = "{QCD quark condensate in external magnetic fields}",
    eprint = "1206.4205",
    archivePrefix = "arXiv",
    primaryClass = "hep-lat",
    doi = "10.1103/PhysRevD.86.071502",
    journal = "Phys. Rev. D",
    volume = "86",
    pages = "071502",
    year = "2012"
}

@article{Bruckmann:2013oba,
    author = "Bruckmann, Falk and Endrodi, Gergely and Kovacs, Tamas G.",
    title = "{Inverse magnetic catalysis and the Polyakov loop}",
    eprint = "1303.3972",
    archivePrefix = "arXiv",
    primaryClass = "hep-lat",
    doi = "10.1007/JHEP04(2013)112",
    journal = "JHEP",
    volume = "04",
    pages = "112",
    year = "2013"
}

@article{Farias:2014eca,
    author = "Farias, R. L. S. and Gomes, K. P. and Krein, G. I. and Pinto, M. B.",
    title = "{Importance of asymptotic freedom for the pseudocritical temperature in magnetized quark matter}",
    eprint = "1404.3931",
    archivePrefix = "arXiv",
    primaryClass = "hep-ph",
    doi = "10.1103/PhysRevC.90.025203",
    journal = "Phys. Rev. C",
    volume = "90",
    number = "2",
    pages = "025203",
    year = "2014"
}

@article{Ferreira:2014kpa,
    author = "Ferreira, M. and Costa, P. and Louren{\c{c}}o, O. and Frederico, T. and Provid{\^e}ncia, C.",
    title = "{Inverse magnetic catalysis in the (2+1)-flavor Nambu-Jona-Lasinio and Polyakov-Nambu-Jona-Lasinio models}",
    eprint = "1404.5577",
    archivePrefix = "arXiv",
    primaryClass = "hep-ph",
    doi = "10.1103/PhysRevD.89.116011",
    journal = "Phys. Rev. D",
    volume = "89",
    number = "11",
    pages = "116011",
    year = "2014"
}

@article{Ayala:2014gwa,
    author = "Ayala, Alejandro and Loewe, M. and Zamora, R.",
    title = "{Inverse magnetic catalysis in the linear sigma model with quarks}",
    eprint = "1406.7408",
    archivePrefix = "arXiv",
    primaryClass = "hep-ph",
    doi = "10.1103/PhysRevD.91.016002",
    journal = "Phys. Rev. D",
    volume = "91",
    number = "1",
    pages = "016002",
    year = "2015"
}

@article{Ayala:2015lta,
    author = "Ayala, Alejandro and Dominguez, C. A. and Hernandez, L. A. and Loewe, M. and Zamora, R.",
    title = "{Magnetized effective QCD phase diagram}",
    eprint = "1509.03345",
    archivePrefix = "arXiv",
    primaryClass = "hep-ph",
    doi = "10.1103/PhysRevD.92.119905",
    journal = "Phys. Rev. D",
    volume = "92",
    number = "9",
    pages = "096011",
    year = "2015",
    note = "[Addendum: Phys.Rev.D 92, 119905 (2015)]"
}

@article{Ayala:2014iba,
    author = "Ayala, Alejandro and Loewe, M. and Mizher, Ana Julia and Zamora, R.",
    title = "{Inverse magnetic catalysis for the chiral transition induced by thermo-magnetic effects on the coupling constant}",
    eprint = "1406.3885",
    archivePrefix = "arXiv",
    primaryClass = "hep-ph",
    doi = "10.1103/PhysRevD.90.036001",
    journal = "Phys. Rev. D",
    volume = "90",
    number = "3",
    pages = "036001",
    year = "2014"
}

@article{Farias:2016gmy,
    author = "Farias, R. L. S. and Timoteo, V. S. and Avancini, S. S. and Pinto, M. B. and Krein, G.",
    title = "{Thermo-magnetic effects in quark matter: Nambu--Jona-Lasinio model constrained by lattice QCD}",
    eprint = "1603.03847",
    archivePrefix = "arXiv",
    primaryClass = "hep-ph",
    doi = "10.1140/epja/i2017-12320-8",
    journal = "Eur. Phys. J. A",
    volume = "53",
    number = "5",
    pages = "101",
    year = "2017"
}

@article{Ayala:2016bbi,
    author = "Ayala, Alejandro and Dominguez, C. A. and Hernandez, L. A. and Loewe, M. and Raya, Alfredo and Rojas, J. C. and Villavicencio, C.",
    title = "{Thermomagnetic properties of the strong coupling in the local Nambu{\textendash}Jona-Lasinio model}",
    eprint = "1603.00833",
    archivePrefix = "arXiv",
    primaryClass = "hep-ph",
    doi = "10.1103/PhysRevD.94.054019",
    journal = "Phys. Rev. D",
    volume = "94",
    number = "5",
    pages = "054019",
    year = "2016"
}

@article{Ferrer:2014qka,
    author = "Ferrer, E. J. and de la Incera, V. and Wen, X. J.",
    title = "{Quark Antiscreening at Strong Magnetic Field and Inverse Magnetic Catalysis}",
    eprint = "1407.3503",
    archivePrefix = "arXiv",
    primaryClass = "nucl-th",
    doi = "10.1103/PhysRevD.91.054006",
    journal = "Phys. Rev. D",
    volume = "91",
    number = "5",
    pages = "054006",
    year = "2015"
}

@article{Ayala:2014uua,
    author = "Ayala, Alejandro and Cobos-Mart{\'\i}nez, J. J. and Loewe, M. and Tejeda-Yeomans, Mar{\'\i}a Elena and Zamora, R.",
    title = "{Finite temperature quark-gluon vertex with a magnetic field in the Hard Thermal Loop approximation}",
    eprint = "1410.6388",
    archivePrefix = "arXiv",
    primaryClass = "hep-ph",
    doi = "10.1103/PhysRevD.91.016007",
    journal = "Phys. Rev. D",
    volume = "91",
    number = "1",
    pages = "016007",
    year = "2015"
}

@article{Ayala:2018wux,
    author = "Ayala, Alejandro and Dominguez, C. A. and Hernandez-Ortiz, Saul and Hernandez, L. A. and Loewe, M. and Manreza Paret, D. and Zamora, R.",
    title = "{Thermomagnetic evolution of the QCD strong coupling}",
    eprint = "1805.08198",
    archivePrefix = "arXiv",
    primaryClass = "hep-ph",
    doi = "10.1103/PhysRevD.98.031501",
    journal = "Phys. Rev. D",
    volume = "98",
    number = "3",
    pages = "031501",
    year = "2018"
}

@article{Mueller:2014tea,
    author = "Mueller, Niklas and Bonnet, Jacqueline A. and Fischer, Christian S.",
    title = "{Dynamical quark mass generation in a strong external magnetic field}",
    eprint = "1401.1647",
    archivePrefix = "arXiv",
    primaryClass = "hep-ph",
    doi = "10.1103/PhysRevD.89.094023",
    journal = "Phys. Rev. D",
    volume = "89",
    number = "9",
    pages = "094023",
    year = "2014"
}

@article{Mueller:2015fka,
    author = "Mueller, Niklas and Pawlowski, Jan M.",
    title = "{Magnetic catalysis and inverse magnetic catalysis in QCD}",
    eprint = "1502.08011",
    archivePrefix = "arXiv",
    primaryClass = "hep-ph",
    doi = "10.1103/PhysRevD.91.116010",
    journal = "Phys. Rev. D",
    volume = "91",
    number = "11",
    pages = "116010",
    year = "2015"
}

@article{Bandyopadhyay:2020zte,
    author = "Bandyopadhyay, Aritra and Farias, Ricardo L. S.",
    title = "{Inverse magnetic catalysis: how much do we know about?}",
    eprint = "2003.11054",
    archivePrefix = "arXiv",
    primaryClass = "hep-ph",
    doi = "10.1140/epjs/s11734-021-00023-1",
    journal = "Eur. Phys. J. ST",
    volume = "230",
    number = "3",
    pages = "719--728",
    year = "2021"
}

@article{Nikishov:1964zza,
    author = "Nikishov, A. I. and Ritus, V. I.",
    title = "{Quantum Processes in the Field of a Plane Electromagnetic Wave and in a Constant Field. I}",
    journal = "Sov. Phys. JETP",
    volume = "19",
    pages = "529--541",
    year = "1964"
}

@article{Nikishov:1964zz,
    author = "Nikishov, A. I. and Ritus, V. I.",
    title = "{Quantum Processes in the Field of a Plane Electromagnetic Wave and in a Constant Field}",
    journal = "Sov. Phys. JETP",
    volume = "19",
    pages = "1191--1199",
    year = "1964"
}

@article{Bali:2018sey,
    author = {Bali, G. S. and Brandt, B. B. and Endr{\H{o}}di, G. and Gl{\"a}{\ss}le, B.},
    title = "{Weak decay of magnetized pions}",
    eprint = "1805.10971",
    archivePrefix = "arXiv",
    primaryClass = "hep-lat",
    doi = "10.1103/PhysRevLett.121.072001",
    journal = "Phys. Rev. Lett.",
    volume = "121",
    number = "7",
    pages = "072001",
    year = "2018"
}

@article{Coppola:2018ygv,
    author = "Coppola, M. and Gomez Dumm, D. and Noguera, S. and Scoccola, N. N.",
    title = "{Pion-to-vacuum vector and axial vector amplitudes and weak decays of pions in a magnetic field}",
    eprint = "1810.08110",
    archivePrefix = "arXiv",
    primaryClass = "hep-ph",
    doi = "10.1103/PhysRevD.99.054031",
    journal = "Phys. Rev. D",
    volume = "99",
    number = "5",
    pages = "054031",
    year = "2019"
}

@article{Coppola:2019idh,
    author = "Coppola, Maximo and Gomez Dumm, Daniel and Noguera, Santiago and Scoccola, Norberto N.",
    title = "{Magnetic field driven enhancement of the weak decay width of charged pions}",
    eprint = "1908.10765",
    archivePrefix = "arXiv",
    primaryClass = "hep-ph",
    doi = "10.1007/JHEP09(2020)058",
    journal = "JHEP",
    volume = "09",
    pages = "058",
    year = "2020"
}

@article{Coppola:2019wvh,
    author = "Coppola, M. and Gomez Dumm, D. and Noguera, S. and Scoccola, N. N.",
    title = "{Weak decays of magnetized charged pions in the symmetric gauge}",
    eprint = "1910.10814",
    archivePrefix = "arXiv",
    primaryClass = "hep-ph",
    doi = "10.1103/PhysRevD.101.034003",
    journal = "Phys. Rev. D",
    volume = "101",
    number = "3",
    pages = "034003",
    year = "2020"
}

@article{Coppola:2025nus,
    author = "Coppola, M{\'a}ximo and Gomez Dumm, Daniel and Scoccola, Norberto N.",
    title = "{{\ensuremath{\pi}}0{\textrightarrow}2{\ensuremath{\gamma}} decay under strong magnetic fields in the NJL model}",
    eprint = "2507.13560",
    archivePrefix = "arXiv",
    primaryClass = "hep-ph",
    doi = "10.1103/3hqb-28zb",
    journal = "Phys. Rev. D",
    volume = "112",
    number = "5",
    pages = "054043",
    year = "2025"
}

@article{Brauner:2017uiu,
    author = "Brauner, Tomas and Kadam, Saurabh V.",
    title = "{Anomalous low-temperature thermodynamics of QCD in strong magnetic fields}",
    eprint = "1706.04514",
    archivePrefix = "arXiv",
    primaryClass = "hep-ph",
    doi = "10.1007/JHEP11(2017)103",
    journal = "JHEP",
    volume = "11",
    pages = "103",
    year = "2017"
}

@misc{Adhikari:2024vhs,
    author = "Adhikari, Prabal and Tiburzi, Brian C.",
    title = "{Chiral Symmetry Breaking and Pion Decay in a Magnetic Field}",
    eprint = "2406.00818",
    archivePrefix = "arXiv",
    primaryClass = "hep-ph",
    month = "6",
    year = "2024"
}

@article{Ayala:2020muk,
    author = "Ayala, Alejandro and Hern{\'a}ndez, Jos{\'e} Luis and Hern{\'a}ndez, L. A. and Farias, Ricardo L. S. and Zamora, R.",
    title = "{Magnetic corrections to the boson self-coupling and boson-fermion coupling in the linear sigma model with quarks}",
    eprint = "2009.13740",
    archivePrefix = "arXiv",
    primaryClass = "hep-ph",
    doi = "10.1103/PhysRevD.102.114038",
    journal = "Phys. Rev. D",
    volume = "102",
    number = "11",
    pages = "114038",
    year = "2020"
}

@article{Ayala:2020dxs,
    author = "Ayala, Alejandro and Hern{\'a}ndez, Jos{\'e} Luis and Hern{\'a}ndez, L. A. and Farias, Ricardo L. S. and Zamora, R.",
    title = "{Magnetic field dependence of the neutral pion mass in the linear sigma model with quarks: The strong field case}",
    eprint = "2011.03673",
    archivePrefix = "arXiv",
    primaryClass = "hep-ph",
    doi = "10.1103/PhysRevD.103.054038",
    journal = "Phys. Rev. D",
    volume = "103",
    number = "5",
    pages = "054038",
    year = "2021"
}

@article{Lin:2021bqv,
    author = "Lin, Fan and Huang, Mei",
    title = "{Magnetic correction to the anomalous magnetic moment of electrons}",
    eprint = "2112.01051",
    archivePrefix = "arXiv",
    primaryClass = "hep-ph",
    doi = "10.1088/1572-9494/ac5b5e",
    journal = "Commun. Theor. Phys.",
    volume = "74",
    number = "5",
    pages = "055202",
    year = "2022"
}

@article{Baier:2000yv,
    author = "Baier, V. N. and Katkov, V. M.",
    title = "{Anomalous magnetic moment of the electron in a medium}",
    eprint = "hep-ph/0011340",
    archivePrefix = "arXiv",
    reportNumber = "BUDKER-INP-2000-91, BUDKER-INP-00-91",
    doi = "10.1016/S0375-9601(01)00077-9",
    journal = "Phys. Lett. A",
    volume = "280",
    pages = "275--281",
    year = "2001"
}

@article{Tavares:2023oln,
    author = "Tavares, William R. and Avancini, Sidney S. and Farias, Ricardo L. S. and Cardoso, Rafael P.",
    title = "{Artificial first-order phase transition in a magnetized Nambu{\textendash}Jona-Lasinio model with a quark anomalous magnetic moment}",
    eprint = "2309.04055",
    archivePrefix = "arXiv",
    primaryClass = "hep-ph",
    doi = "10.1103/PhysRevD.109.016011",
    journal = "Phys. Rev. D",
    volume = "109",
    number = "1",
    pages = "016011",
    year = "2024"
}

@article{Farias:2021fci,
    author = "Farias, Ricardo L. S. and Tavares, William R. and Nunes, Rodrigo M. and Avancini, Sidney S.",
    title = "{Effects of the quark anomalous magnetic moment in the chiral symmetry restoration: magnetic catalysis and inverse magnetic catalysis}",
    eprint = "2109.11112",
    archivePrefix = "arXiv",
    primaryClass = "hep-ph",
    doi = "10.1140/epjc/s10052-022-10640-2",
    journal = "Eur. Phys. J. C",
    volume = "82",
    number = "8",
    pages = "674",
    year = "2022"
}

@article{Fayazbakhsh:2014mca,
    author = "Fayazbakhsh, Sh. and Sadooghi, N.",
    title = "{Anomalous magnetic moment of hot quarks, inverse magnetic catalysis, and reentrance of the chiral symmetry broken phase}",
    eprint = "1408.5457",
    archivePrefix = "arXiv",
    primaryClass = "hep-ph",
    doi = "10.1103/PhysRevD.90.105030",
    journal = "Phys. Rev. D",
    volume = "90",
    number = "10",
    pages = "105030",
    year = "2014"
}

@misc{Ayala:2026eja,
  author = {Ayala, Alejandro and Mu{\~n}oz, Enrique and Loewe, Marcelo and Rojas, Juan Cristobal and Scoccola, Norberto},
  title = {\mbox{QED} Vertex and Anomalous Magnetic Moment in the Presence of a Magnetic Field},
  year = {2026},
  eprint = {2607.10015},
  archivePrefix = {arXiv},
  primaryClass = {hep-ph}
}

@article{Fraga:2024klm,
    author = "Fraga, Eduardo S. and Palhares, Leticia F. and Villavicencio, Cristian",
    title = "{Quark anomalous magnetic moment in an extreme magnetic background from perturbative QCD}",
    eprint = "2403.10641",
    archivePrefix = "arXiv",
    primaryClass = "hep-ph",
    doi = "10.1103/PhysRevD.109.116018",
    journal = "Phys. Rev. D",
    volume = "109",
    number = "11",
    pages = "116018",
    year = "2024"
}

@article{GomezDumm:2023owj,
    author = "Gomez Dumm, D. and Noguera, S. and Scoccola, N. N.",
    title = "{Charged meson masses under strong magnetic fields: Gauge invariance and Schwinger phases}",
    eprint = "2306.04128",
    archivePrefix = "arXiv",
    primaryClass = "hep-ph",
    doi = "10.1103/PhysRevD.108.016012",
    journal = "Phys. Rev. D",
    volume = "108",
    number = "1",
    pages = "016012",
    year = "2023"
}

@article{Ritus:1978cj,
    author = "Ritus, V. I.",
    title = "{METHOD OF EIGENFUNCTIONS AND MASS OPERATOR IN QUANTUM ELECTRODYNAMICS OF A CONSTANT FIELD}",
    reportNumber = "LEBEDEV-78-242",
    journal = "Sov. Phys. JETP",
    volume = "48",
    pages = "788",
    year = "1978"
}
\bibliographystyle{apsrev4-1}

\end{document}